\documentclass[useAMS,usenatbib]{mn2e}

\usepackage{graphicx}
\usepackage{amsmath}
\usepackage{amssymb}
\usepackage{color}
\usepackage{subfig}
\usepackage[breaklinks,colorlinks,citecolor=blue,linkcolor=red]{hyperref} 
\usepackage[all]{hypcap}

\newcommand\msun{\, \rm M_\odot}

\newcommand\kms{\, \rm km\,s^{-1}}

\newcommand\gyr{{\, \rm Gyr}}

\newcommand\eout{{e_{\rm out}}}
\newcommand\ein{{e_{\rm in}}}
\newcommand\aout{{a_{\rm out}}}
\newcommand\ain{{a_{\rm in}}}
\newcommand\amax{{a_{\rm 3, max}}}
\newcommand\mbh{{m_{\rm BH}}}
\newcommand\mns{{m_{\rm NS}}}
\newcommand\sigbh{{\sigma_{\rm BH}}}
\newcommand\signs{{\sigma_{\rm NS}}}
\newcommand\chibh{{\chi_{\rm BH}}}
\newcommand\chins{{\chi_{\rm NS}}}
\newcommand\chieff{{\chi_{\rm eff}}}
\newcommand\thetabh{{\theta_{\rm BH}}}
\newcommand\thetans{{\theta_{\rm NS}}}

\newcommand\vk{{v_\mathrm{k}}}

\title[BH-NS mergers from triples II]{Black hole-neutron star mergers from triples II: the role of metallicity and spin-orbit misalignment}
\author[G. Fragione, A. Loeb]{\parbox{\textwidth}{Giacomo Fragione$^{1,2}$\thanks{E-mail: giacomo.fragione@northwestern.edu}, Abraham Loeb$^{3}$}\\
$^1$Department of Physics \& Astronomy, Northwestern University, Evanston, IL 60202, USA\\
$^2$Center for Interdisciplinary Exploration \& Research in Astrophysics (CIERA), Evanston, IL 60202, USA\\
$^3$Astronomy Department, Harvard University, 60 Garden St., Cambridge, MA 02138, USA}

\hypersetup{draft}

\begin{document}

\maketitle

\begin{abstract}
Observations of black hole-neutron star (BH-NS) mergers via gravitational waves (GWs) are of great interest for their electromagnetic counterparts, such as short gamma-ray bursts, and could provide crucial information on the nature of BHs and the NS crust and magnetosphere. While no event has been confirmed, a recent possible detection of a BH-NS merger event by the LIGO-Virgo collaboration has attracted a lot of attention to these sources. In this second paper of the series, we follow-up our study of the dynamical evolution of triples comprised of an inner BH-NS binary. In particular, we examine how the progenitor metallicity affects the characteristics of the BH-NS mergers in triples. We determine the distributions of masses, orbital parameters and merger times, as a function of the progenitor metallicity and initial triple orbital distributions, and show that the typical eccentricity in the LIGO band is $\sim 10^{-2}-10^{-1}$. We derive a merger rate range of $\Gamma_\mathrm{BH-NS}=1.9\times 10^{-4}-22 \ \mathrm{Gpc}^{-3}\ \mathrm{yr}^{-1}$, consistent the LIGO-Virgo upper limit. Finally, we study the expected spin-orbit misalignments of merging BH-NS binaries from this channel, and find that typically the effective spin distribution is peaked at $\chieff\sim 0$ with significant tails. 
\end{abstract}


\section{Introduction}

Several different astrophysical mechanisms have been proposed to explain the black hole (BH) and neutron star (NS) binary mergers observed via gravitational wave (GW) emission by the LIGO-Virgo collaboration. Scenarios include isolated binary evolution \citep{bel16b,demink2016,kruc2018,eldridge2019}, mergers in star clusters \citep{askar17,baner18,frak18,rod18}, GW capture events and Kozai-Lidov (KL) mergers in galactic nuclei \citep{olea09,antoper12,fragrish2018,grish18,rass2019}, mergers in active galactic nuclei accretion disks \citep{bart17}, and KL mergers in isolated triple and quadruple systems \citep{ant17,sil17,arc2018,fragk2019,liu2019}.

\begin{table*}
\caption{Description of important quantities used in the text.}
\centering
\begin{tabular}{lc}
\hline
Symbol & Description\\
\hline\hline
$\mbh$ & Mass of the black hole in the inner binary \\
$\mns$ & Mass of the neutron star in the inner binary \\
$m_3$ & Mass of the third companion in the triple \\
$\ain$ & Semi-major axis of the inner (black hole-neutron star) orbit \\
$\ein$ & Eccentricity of the inner (black hole-neutron star) orbit \\
$\aout$ & Semi-major axis of the outer orbit \\
$\eout$ & Eccentricity of the outer orbit \\
$i_0$ & Inclination between the inner and outer orbits \\
$\amax$ & Maximum outer semi-major axis of the triple \\
$\sigbh$ & Dispersion of black hole kick-velocity distribution \\
$\signs$ & Dispersion of neutron star kick-velocity distribution \\
$Z$ & Progenitor metallicity \\
$\chibh$ & Kerr parameter of the black hole \\
$\chins$ & Kerr parameter of the neutron star \\
\hline
\end{tabular}
\label{tab:quant}
\end{table*}

LIGO-Virgo promise to observe a large number of BH-BH, BH-NS and NS-NS mergers in the next few years and to shed light on their origin. Thus, it is of crucial importance to examine the different contributions to the overall observed rate. It has been shown that the distributions of masses, spins, eccentricity and redshift of the merging compact objects could be used to statistically disentangle the contributions of different origins \citep[see e.g.][]{olea16,gondan2018}. For example, BHs and NSs merging in triples and quadruples are expected to retain significant eccentricities when entering the LIGO band ($10$ Hz), much larger than binaries that merge in isolation \citep[see e.g.][]{antchrod2016,fragrish2018,frbr2019}.

The O2 catalogue of compact object mergers released by the LIGO-Virgo collaboration includes ten BH binaries and one NS binary \citep{ligo2018}. No BH-NS mergers have been observed, even though there are now possible candidate events amongst GW detections (see e.g. LVC GCN 24237 and LVC GCN 25324). BH-NS mergers might have electromagnetic (EM) counterparts, such as short gamma-ray bursts, which can provide crucial information on the related accretion onto stellar BHs and provide unique information on the NS crust and magnetosphere \citep{Pannarale2011,Foucart2012,tsang2012,dora2013}.

The origin of BH-NS mergers is still highly uncertain and debated. BH-NS binaries can be produced in isolation as a result of binary evolution, as for BH-BH and NS-NS binaries \citep{demink2016,kruc2018,eldridge2019}. More complicated is the process of forming BH-NS binaries through dynamical assembly in star clusters. A number of papers showed that NSs are generally prevented from forming NS-NS and BH-NS binaries in a star cluster as a result of the strong heating due to gravitational BH scatterings \citep{frag2018,ye2019}. Only if most of the BHs have been ejected from the cluster, NSs can efficiently segregate in the innermost regions and possibly form binaries, that later merge. Recently, \citet{frl2019} have proposed that BH-NS mergers can be a natural outcome of the dynamical evolution of triple systems. Here, the KL cycles imposed by the tidal field of the tertiary can make the inner BH-NS binary reach high eccentricities and merge as a consequence of the efficient dissipation of energy through GW emission at pericentre. 

In this second paper of the series, we follow-up our study of the dynamical evolution of triples comprised of an inner BH-NS binary \citep{frl2019}, by means of high-precision $N$-body simulations, including Post-Newtonian (PN) terms up to 2.5PN. We start from the main sequence progenitors of the BHs and model the supernova (SN) events that lead to the formation of the BH triple. We adopt different prescriptions for SN natal kicks and orbital parameters. We also take into account different progenitor metallicities to study how this parameter affects the characteristics of the BH-NS mergers in triples. We determine the distributions of various merger properties, including masses, eccentricities, and merger times. Finally, we also investigate the expected spin-orbit misalignments of merging BH-NS binaries from this channel.

The paper is organized as follows. In Section~\ref{sect:method}, we discuss the initial conditions adopted in this paper. In Section~\ref{sect:results}, we discuss the parameters of merging systems and present the distributions of masses, eccentricities and spin-orbit misalignments. Finally, in Section~\ref{sect:conc}, we discuss the implications of our findings and summarize our conclusions.

\section{Method}
\label{sect:method}

\begin{table*}
\caption{Models parameters: name, dispersion of BH kick-velocity distribution ($\sigbh$), dispersion of the NS kick-velocity distribution ($\signs$), progenitor metallicity ($Z$), maximum outer semi-major axis of the triple ($\amax$), fraction of stable triple systems after SNe ($f_{\rm stable}$), fraction of stable systems that merge from the $N$-body simulations ($f_{\rm merge}$).}
\centering
\begin{tabular}{lcccccc}
\hline
Name & $\sigbh$ ($\kms$) & $\signs$ ($\kms$) & $Z$ & $\amax$ (AU) & $f_{\rm stable}$ & $f_{\rm merge}$\\
\hline\hline
A1 & $\signs\times (m_\mathrm{NS}/m_\mathrm{BH})$ & $260$ & $0.01$   & $2000$ & $2.6\times 10^{-7}$ & $0.13$\\
A2 & $\signs\times (m_\mathrm{NS}/m_\mathrm{BH})$ & $100$ & $0.01$   & $2000$ & $1.8\times 10^{-5}$ & $0.11$\\
A3 & $\signs\times (m_\mathrm{NS}/m_\mathrm{BH})$ & $0$   & $0.01$   & $2000$ & $1.4\times 10^{-2}$ & $0.13$\\
B1 & $\signs\times (m_\mathrm{NS}/m_\mathrm{BH})$ & $260$ & $0.0001$ & $2000$ & $5.5\times 10^{-4}$ & $0.10$\\
B2 & $\signs\times (m_\mathrm{NS}/m_\mathrm{BH})$ & $260$ & $0.001$  & $2000$ & $1.4\times 10^{-4}$ & $0.07$\\
B3 & $\signs\times (m_\mathrm{NS}/m_\mathrm{BH})$ & $260$ & $0.005$  & $2000$ & $1.6\times 10^{-6}$ & $0.11$\\
B4 & $\signs\times (m_\mathrm{NS}/m_\mathrm{BH})$ & $260$ & $0.015$  & $2000$ & $5.1\times 10^{-8}$ & $0.12$\\
C1 & $\signs\times (m_\mathrm{NS}/m_\mathrm{BH})$ & $260$ & $0.01$   & $5000$ & $1.4\times 10^{-7}$ & $0.11$\\
\hline
\end{tabular}
\label{tab:models}
\end{table*}

First, we describe our initial population of stellar triples. In total, we consider eight different models (see Table~\ref{tab:models}).

We consider a triple system that consists of an inner binary of mass $m_{\rm in}=m_1+m_2$ ($m_1>m_2$) and a third body of mass $m_3$ that orbits the inner binary. The semi-major axis and eccentricity of the inner orbit are $\ain$ and $\ein$, respectively, while the semi-major axis and eccentricity of the outer orbit are $\aout$ and $\eout$, respectively. The inner and outer orbital plane have initial mutual inclination $i_0$. 

In all our models, we sample the mass $m_1$ from a canonical initial mass function \citep{kroupa2001},
\begin{equation}
\frac{dN}{dm} \propto m^{-2.3}\ ,
\label{eqn:bhmassfunc}
\end{equation}
in the mass range $20\msun$-$150\msun$, reflecting the progenitor of the BH. We adopt a flat mass ratio distribution for both the inner binary, $m_2/m_1$, and the outer binary, $m_3/(m_1+m_2)$ \citep{sana12,duch2013,sana2017}. We sample the mass of the secondary in the inner binary in the mass range $8\msun$-$20\msun$, reflecting the progenitor of the NS. The mass third star is sampled in the range $0.5\msun$-$150\msun$\footnote{We are not taking into account a possible dependence of the IMF on the metallicity \citep{marks2012}}.

We sample the distribution of the inner and outer semi-major axis, $\ain$ and $\aout$, respectively, from a log-uniform \citep{kob2014}. We set a minimum inner separation of $10$ AU \citep{frl2019}, while we adopt two different values for the maximum separation of the triple, $\amax=2000$ AU--$5000$ AU \citep{sana2014}. We then assume a flat distribution for the orbital eccentricities of the inner binary, $\ein$, and outer binary, $\eout$ \citep{ant17}. Finally, the initial mutual inclination $i_0$ between the inner and outer orbits is drawn from an isotropic distribution, while the other relevant angles are drawn randomly.

After sampling the relevant parameters, we check that the initial configuration satisfies the stability criterion of hierarchical triples of \citet{mar01}. Otherwise, we sample the triple parameters again according to the above procedure.

We assume that the stars in the inner binary undergo a SN event sequentially. We also assume that every SN takes place instantaneously, that is on a time-scale shorter than the orbital period, during which the exploding star has an instantaneous removal of mass and is converted to a BH ($m_1$) or a NS ($m_2$) \citep{pijloo2012,toonen2016,frl2019}. We determine the final mass $\mbh$ of the BH by using the fitting formulae to the results of the \textsc{parsec} stellar evolution code \citep[see Appendix C in][]{spera2015}. We adopt five different values of the metallicity, $Z=0.0001$--$0.001$--$0.005$--$0.01$--$0.015$, which ultimately set the final mass of the BH remnant\footnote{We ignore the SN-shell impact on the companion stars. We are not modelling the mass loss during neither possible episodes of Roche-lobe overflows nor possible common evolution phases. Both these processes are not well understood and modeled in triple systems. For recent discussion see \citet{rosa2019} and \citet{hamd2019}.}. The final mass of the secondary, which produces a NS, is set to $\mns=1.3\msun$.

As a result of the mass loss, the exploding star is imparted a kick to its center of mass \citep{bla1961}. Furthermore, the system receives a natal kick due to recoil from an asymmetric supernova explosion. We assume that the natal velocity kick is drawn from a Maxwellian distribution,
\begin{equation}
p(\vk)\propto \vk^2 e^{-\vk^2/\sigma^2}\ ,
\label{eqn:vkick}
\end{equation}
with a mean velocity $\sigma$. The value of $\sigma$ is highly uncertain. We implement momentum-conserving kicks, i.e. we assume that the momentum imparted to a BH is the same as the momentum given to a NS \citep{fryer2001}. Therefore, the natal kick velocity for the BHs is simply $\sigbh=\signs\times (1.3\msun/\mbh)=\sigma\times (1.3\msun/\mbh)$. In our fiducial model, we consider $\sigma=260 \kms$ \citep{hobbs2005}. Additionally, we run a model where we set $\sigma=100 \kms$ \citep{arz2002}, and also adopt a model where no natal kick (i.e. $\sigma=0\kms$) is imparted during BH and NS formation. For NSs, this could reflect the formation process of electron-capture SN \citep{pod2004}. In the case $m_3\ge 8\msun$, we let it undergo an SN event and conversion to a compact object as well. In the case it collapses to a BH, the final mass $m_3^{\rm fin}$ is computed using the same formulae used for the primary star, while $m_3^{\rm fin}=1.3\msun$ in the case it produces a NS. If $m_3< 8\msun$, then $m_3^{\rm fin}=m_3$.

After each SN event, the orbital elements of the triple are updated as appropriate \citep[see e.g.][]{frl2019}, to account both for mass loss and natal kicks \citep{bla1961}. We also check again that the stability criterion of hierarchical triples of \citet{mar01} is satisfied and the triple is stable. After all the SNe take place, we integrate the triple systems by means of the \textsc{ARCHAIN} code \citep{mik06,mik08}, a fully regularized code able to model the evolution of binaries of arbitrary mass ratios and eccentricities with high accuracy and that includes PN corrections up to order PN2.5. We performed $\sim 700$--$1000$ simulations for each model in Table~\ref{tab:models}. We fix the maximum integration time as \citep{sil17,frl2019},
\begin{equation}
T=\min \left(10^3 \times T_{\rm KL}, 10\ \gyr \right)\ ,
\label{eqn:tint}
\end{equation}
where $T_{\rm KL}$ is the triple KL timescales,
\begin{equation}
T_{\rm KL}=\frac{8}{15\pi}\frac{m_{\rm tot}}{m_3^{\rm fin}}\frac{P_{\rm out}^2}{P_{\rm in}}\left(1-e_{\rm out}^2\right)^{3/2}\ .
\end{equation}
Here, $m_{\rm tot}=\mbh+\mns+m_3^{\rm fin}$ and $P_{\rm in}$ and $P_{\rm out}$ are the inner and outer orbital period, respectively. In the case the third companion does not collapse to a compact object, i.e. $m_3< 8\msun$, we set as the maximum timescale the minimum between Eq.~\ref{eqn:tint} and its MS lifetime, which is simply parametrised as \citep[e.g.][]{iben91,hurley00,maeder09},
\begin{equation}
\tau_{\rm MS} = \max(10\ (m/\msun)^{-2.5}\,{\rm Gyr}, 7\,{\rm Myr})\ .
\end{equation}
In this case, we also check if the third star overflows its Roche lobe \citep{egg83}. In such a case, we stop the integration\footnote{We do not model the process that leads to the formation of a white dwarf for the third companion. If the tertiary becomes a white dwarf and the system remains bound, some of the systems could still merge via KL oscillations.}.

\section{Results}
\label{sect:results}

Next, we discuss the parameters of merging BH-NS systems and present the distributions of masses, eccentricities and spin-orbit misalignments. Finally, we also compute the typical merger rate of BH-NS systems in triples.

\begin{figure*} 
\centering
\includegraphics[scale=0.55]{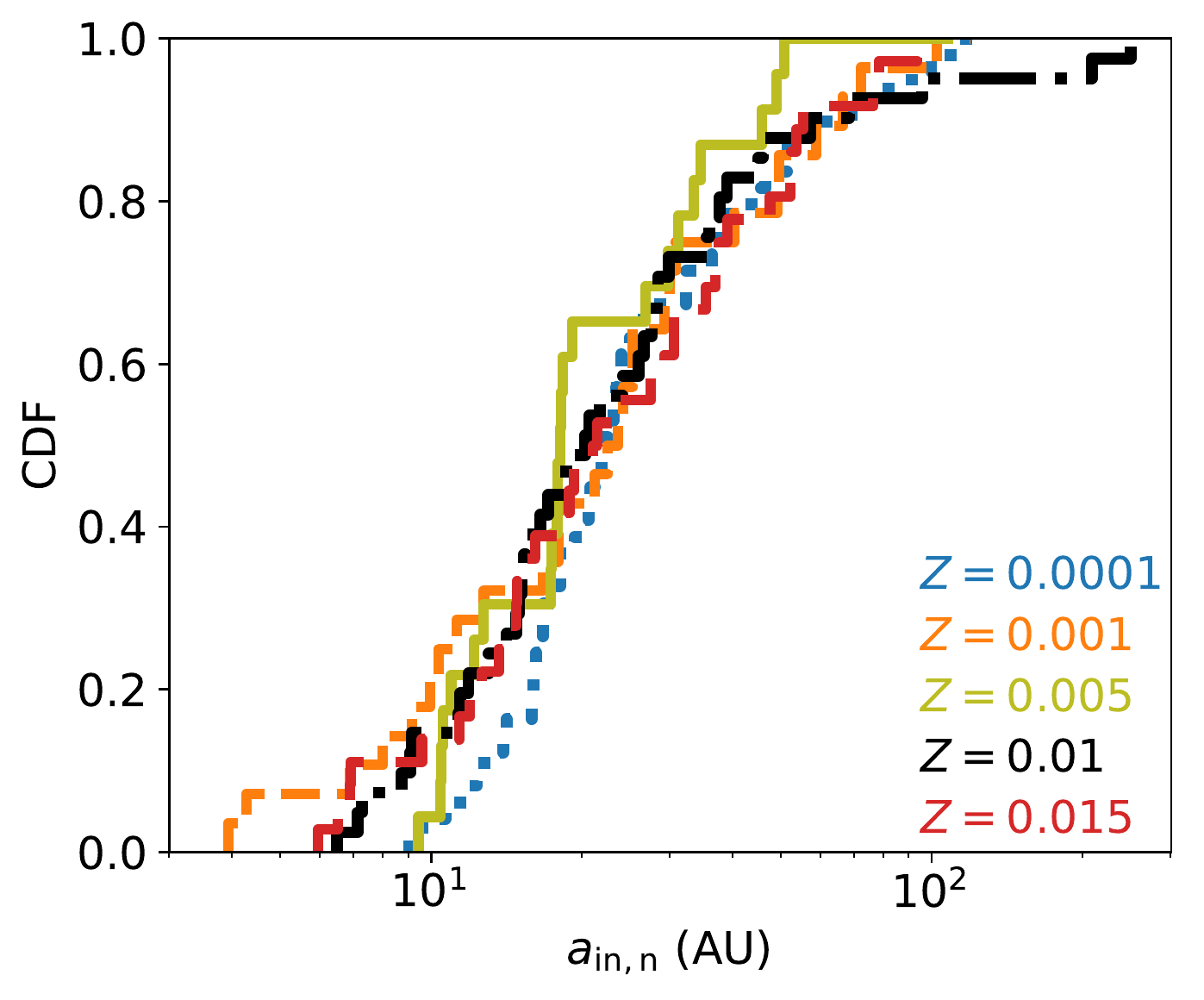}
\hspace{0.5cm}
\includegraphics[scale=0.55]{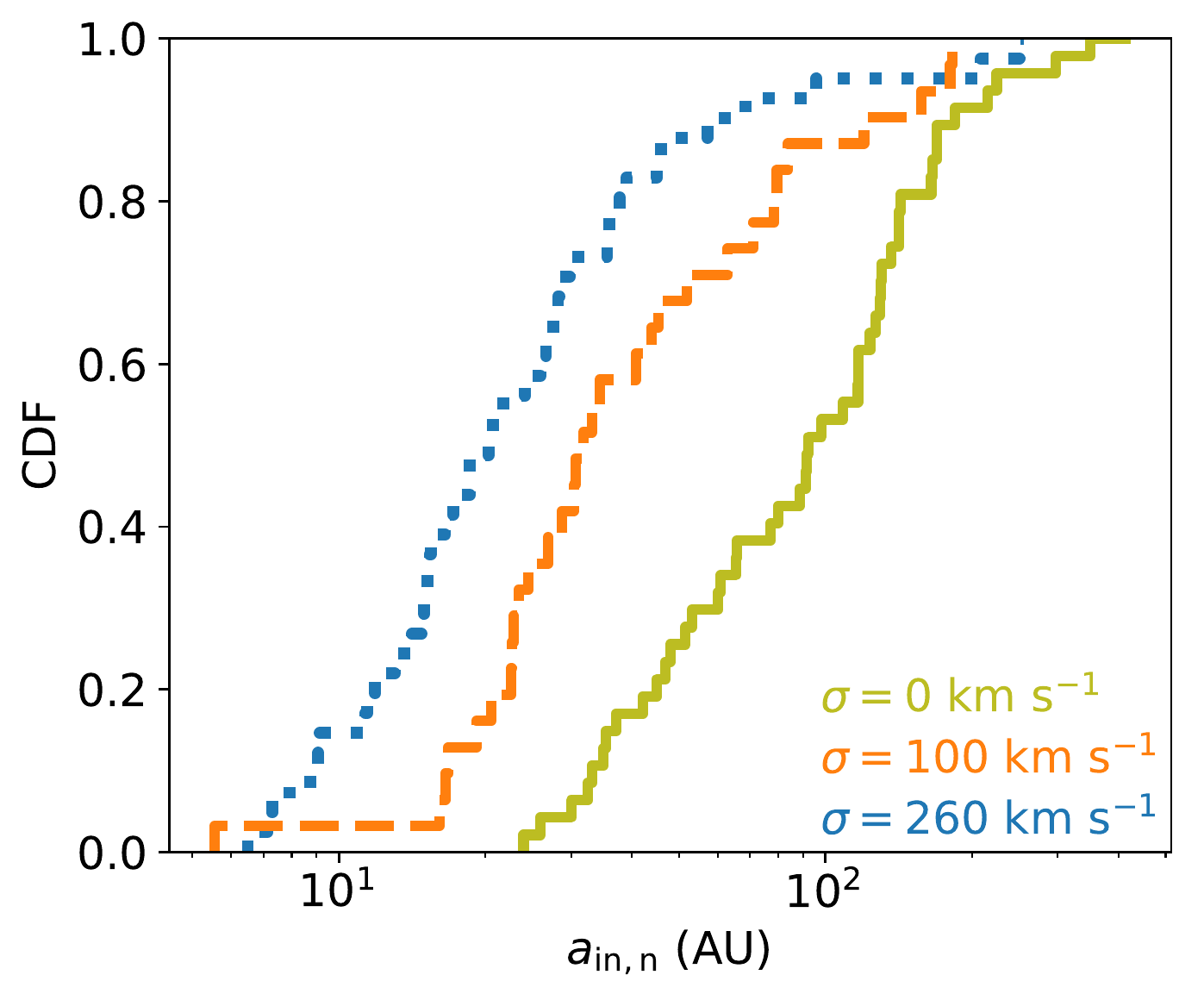}\\
\includegraphics[scale=0.55]{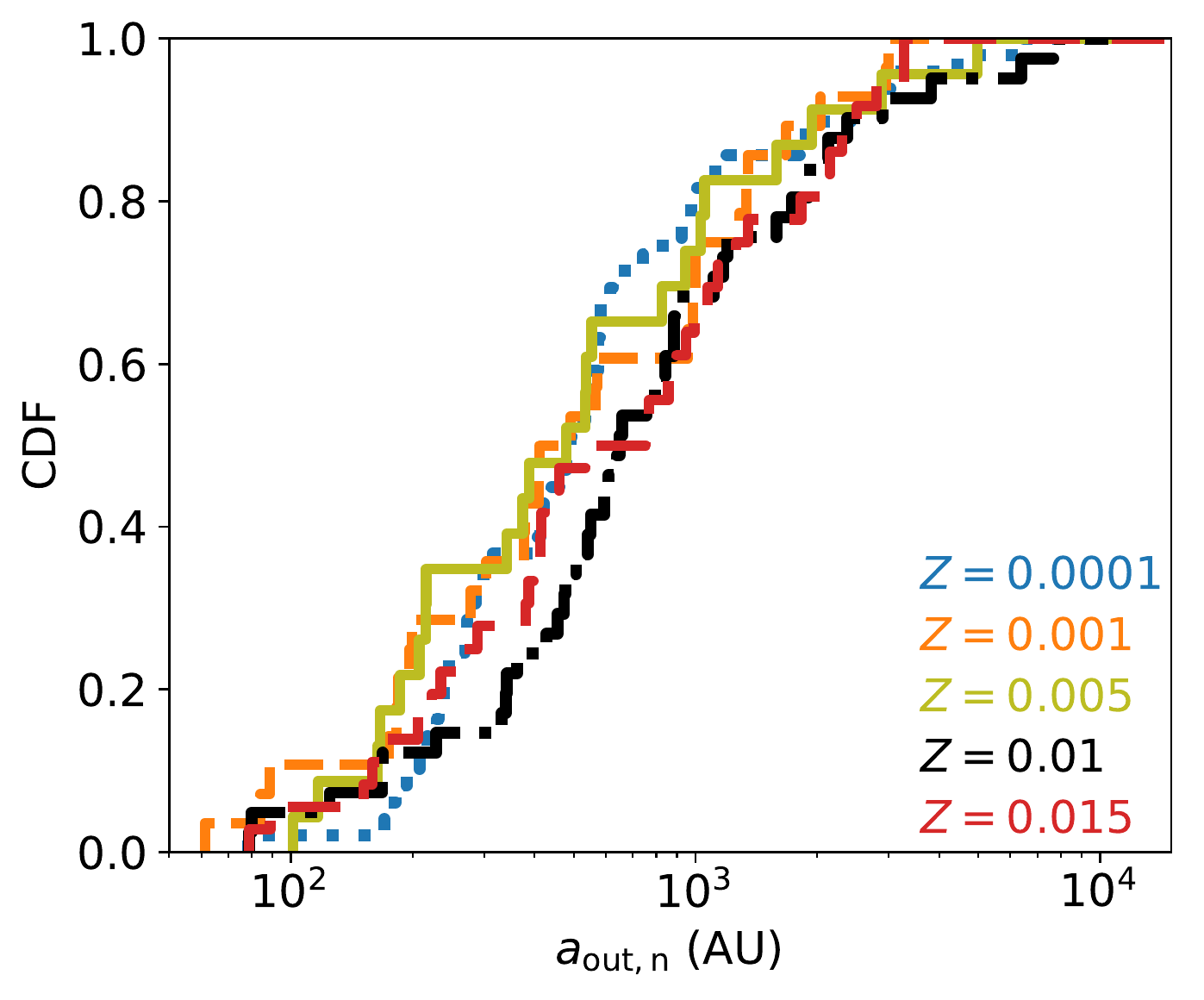}
\hspace{0.5cm}
\includegraphics[scale=0.55]{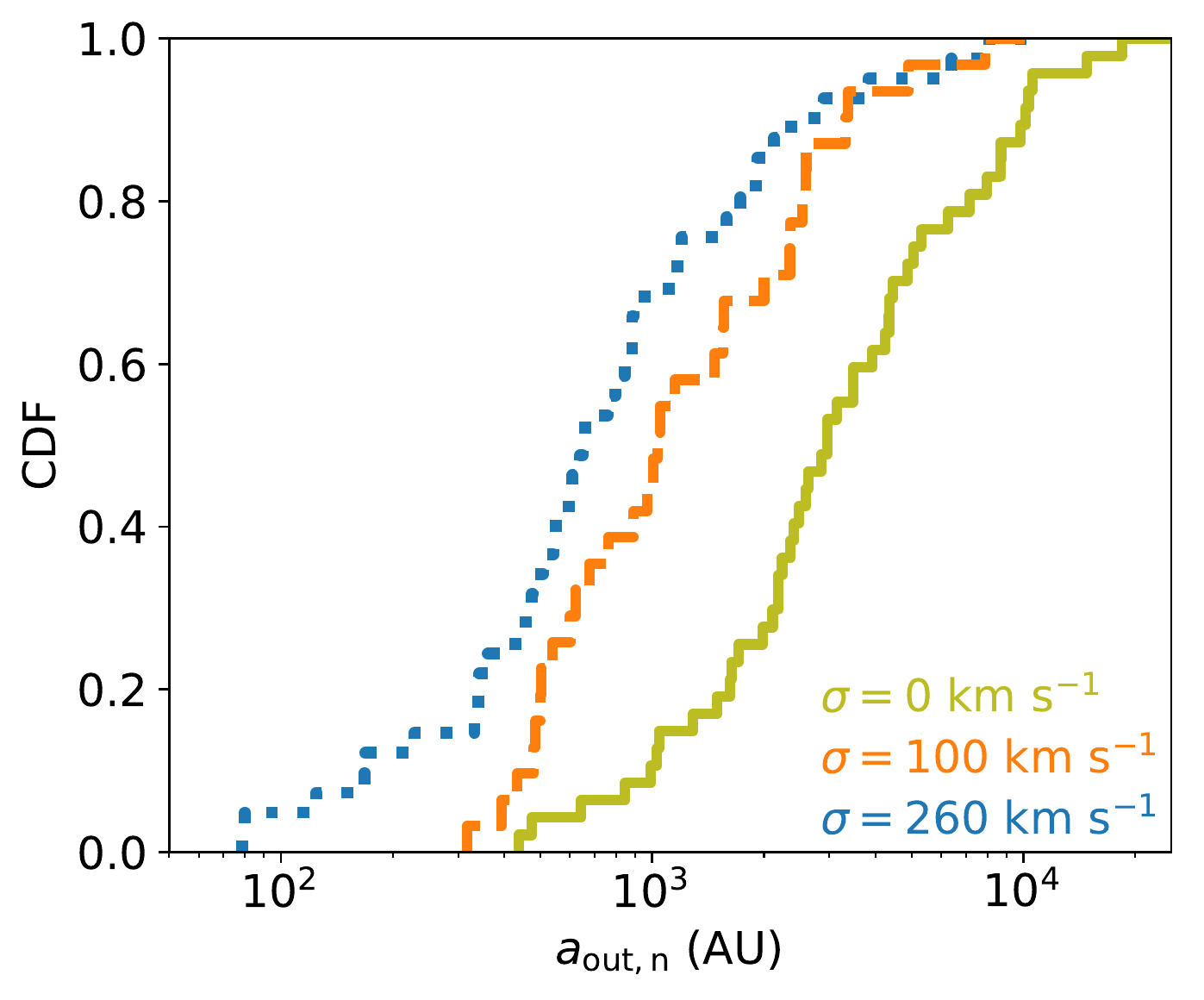}
\caption{Cumulative distribution function of inner (top) and outer (bottom) semi-major axis of BH-NS binaries in triples that lead to a merger. Left panel: $\ain$ and $\aout$ for different progenitor metallicities $Z$ ($\sigma=260\kms$ and $\amax=2000$ AU); right panel: $\ain$ and $\aout$ for different values of $\sigma$ ($Z=0.01$ and $\amax=2000$ AU).}
\label{fig:ainaout}
\end{figure*}

\subsection{Inner and outer semi-major axis}

Figure~\ref{fig:ainaout} reports the cumulative distribution function (CDF) of inner (top) and outer (bottom) semi-major axis of BH-NS binaries in triples that lead to a merger. In the left panel, we show the distribution of $\ain$ and $\aout$ for different values of the metallicity $Z$ ($\sigma=260\kms$ and $\amax=2000$ AU). We find that the metallicity does not affect the distribution of the inner and outer semi-major axes. In right panel, we show the CDFs of $\ain$ and $\aout$ for different values of $\sigma$, for Models A1-A2-A3. In these models, $Z=0.01$ and $\amax=2000$ AU. As shown in \citet{frl2019}, the value of the mean natal velocity kick of Eq.~\ref{eqn:vkick} affects the distribution of the semi-major axes of the triples that lead to a BH-NS merger. We find that $\sim 50$\% of the systems have $a_{\rm in}\lesssim 100$ AU, $\lesssim 30$ AU, $\lesssim 20$ AU for $\sigma=0\kms$, $100\kms$, $260\kms$, respectively, and $\sim 50$\% of the systems have $a_{\rm out}\lesssim 3000$ AU, $\lesssim 1000$ AU, $\lesssim 700$ AU for $\sigma=0\kms$, $100\kms$, $260\kms$, respectively. 

\begin{figure} 
\centering
\includegraphics[scale=0.55]{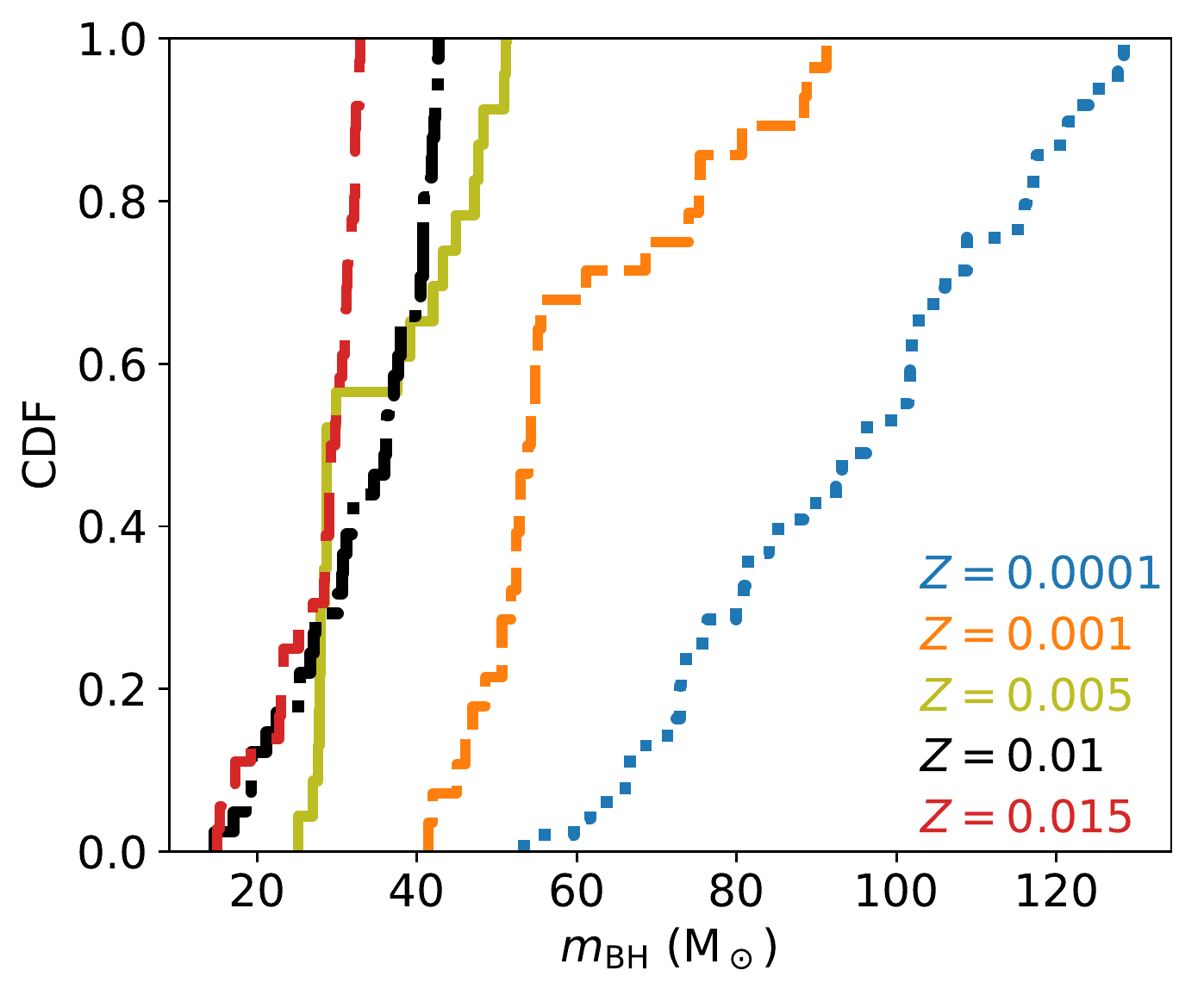}
\includegraphics[scale=0.55]{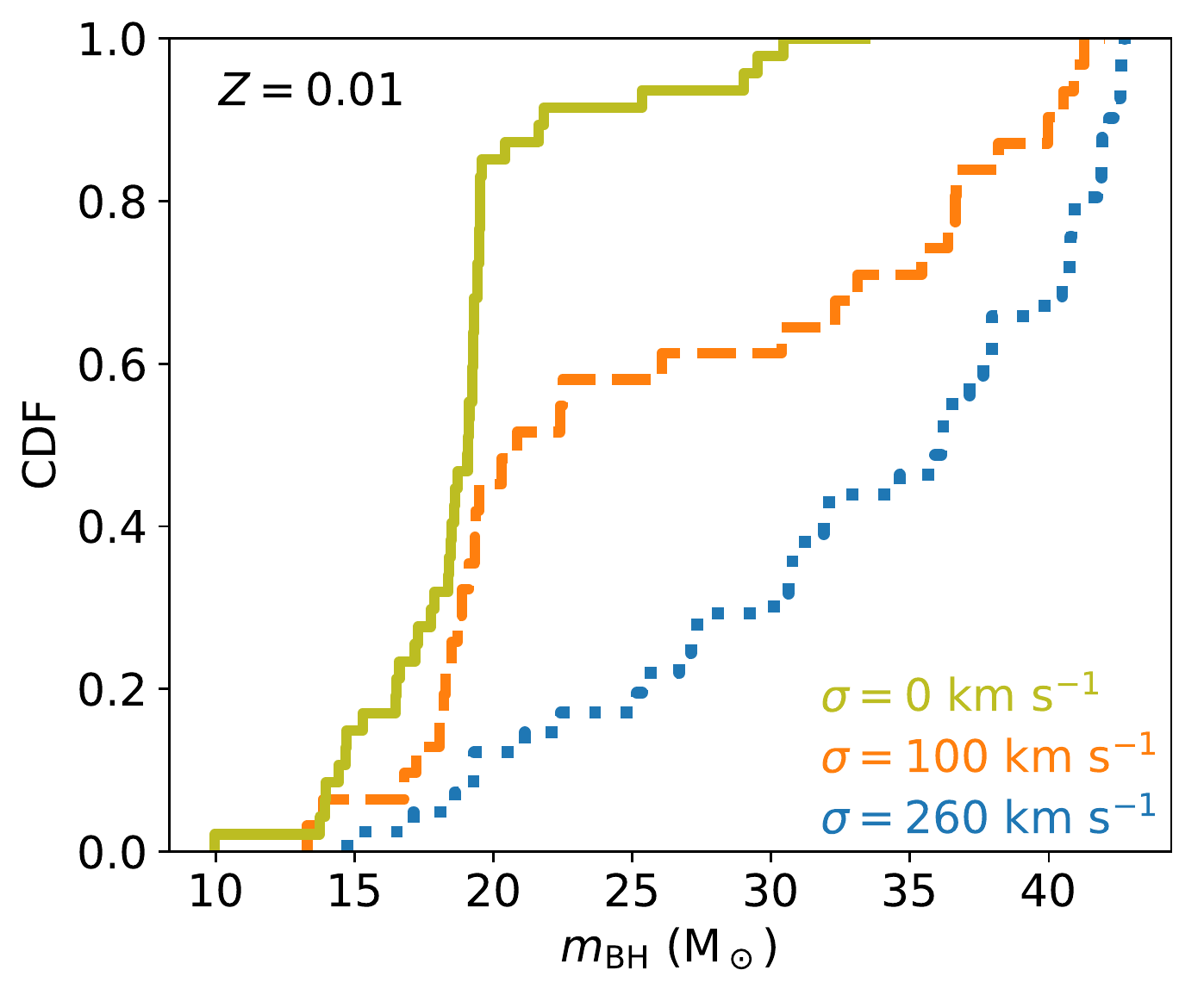}
\caption{Cumulative distribution function of the BH mass of BH-NS binaries in triples that lead to a merger, for different values of $Z$ (top) and $\sigma$ (bottom).}
\label{fig:mass}
\end{figure}

\subsection{Mass distribution}

The typical values of $Z$ and $\sigma$ are expected to have an impact on the distribution of BH masses. Lower metallicity progenitors collapse to make heavier BHs. For $Z\gtrsim 0.002$, the final BH mass is limited to $\sim 60\msun$, while for smaller metallicities the progenitor can even collapse to make a BH of mass $\sim 120\msun$--$140\msun$ \citep{spera2015}. The effect of $\sigma$ on the BH mass distribution can be understood in terms of our assumption of momentum-conserving kicks, where higher mass BHs receive lower velocity kicks, since $\sigbh\propto m_{\rm BH}^{-1}$. Therefore, more massive BHs, which are preferentially produced from low-metallicity progenitors, are more likely to be retained in triples and eventually merge with the NS. 

We illustrate how the CDF of $\mbh$ of BH-NS binaries in triples that lead to a merger depends on the progenitor metallicity in the top panel of Figure~\ref{fig:mass}, for $\sigma=260\kms$ and $\amax=2000$ AU. For $Z\gtrsim 0.005$, we find that the distribution of BH masses is not significantly affected by value of the progenitor metallicity. For these values of $Z$, we find that most of the mergers have $\mbh\lesssim 40\msun$. On the other hand, $\mbh$ is in the range $\sim 40\msun-80\msun$ and $\sim 60\msun-130\msun$ for $Z=0.001$ and $Z=0.0001$, respectively. 

In the bottom panel of Figure~\ref{fig:mass}, we illustrate how the CDF of $\mbh$ of BH-NS binaries in triples that lead to a merger depends on the mean natal kick, for $Z=0.01$ and $\amax=2000$ AU. In the case of $\sigma=0\kms$, we find that $\sim 50\%$ of the BHs that merge have mass $\lesssim 18 \msun$, while $\sim 50\%$ of the BHs that merge have mass $\lesssim 22 \msun$ and $\lesssim 37 \msun$ for $\sigma=100\kms$ and $\sigma=260\kms$, respectively\footnote{Other models for the kicks may lead to different mass distributions.}.

\begin{figure} 
\centering
\includegraphics[scale=0.55]{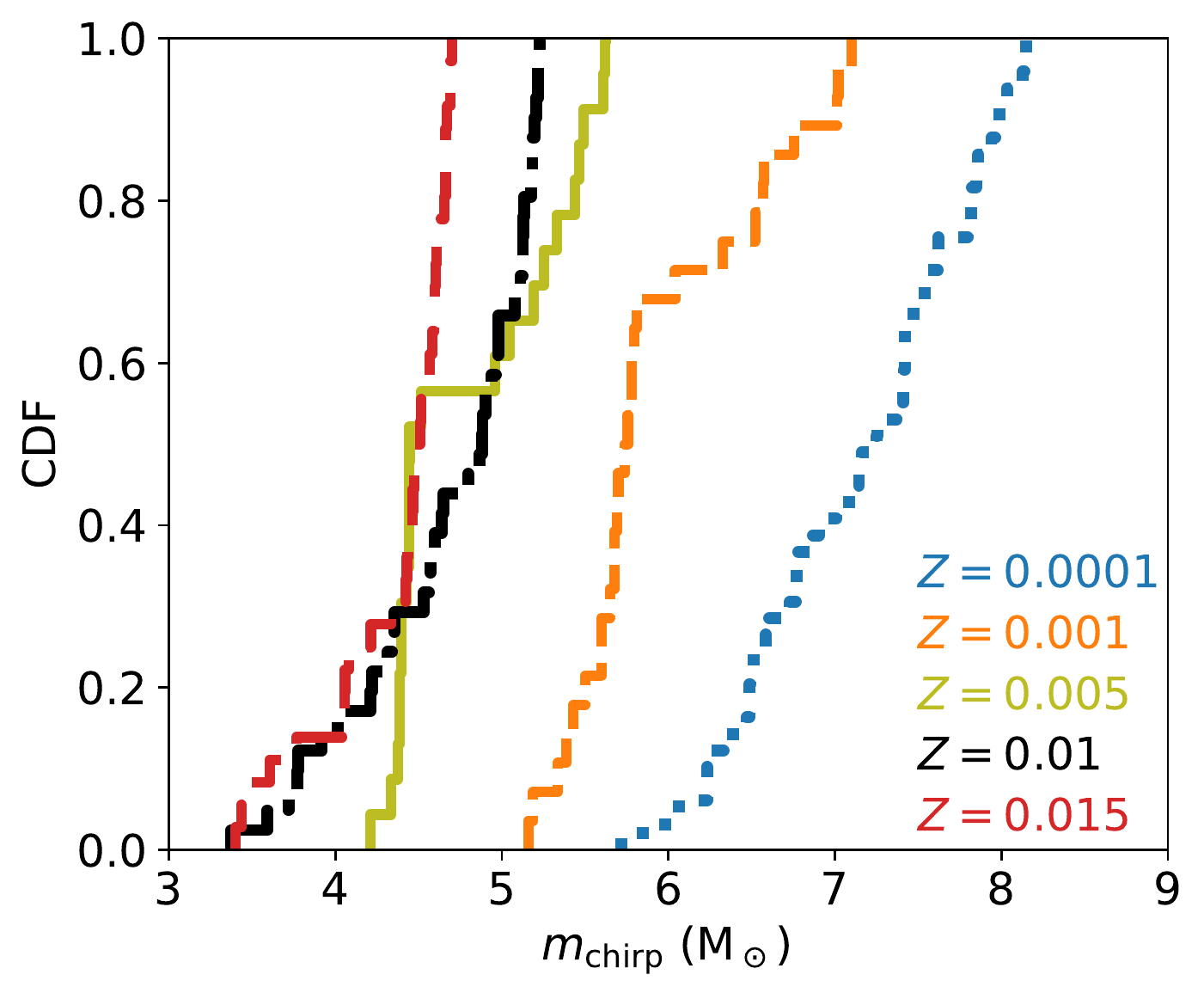}
\includegraphics[scale=0.55]{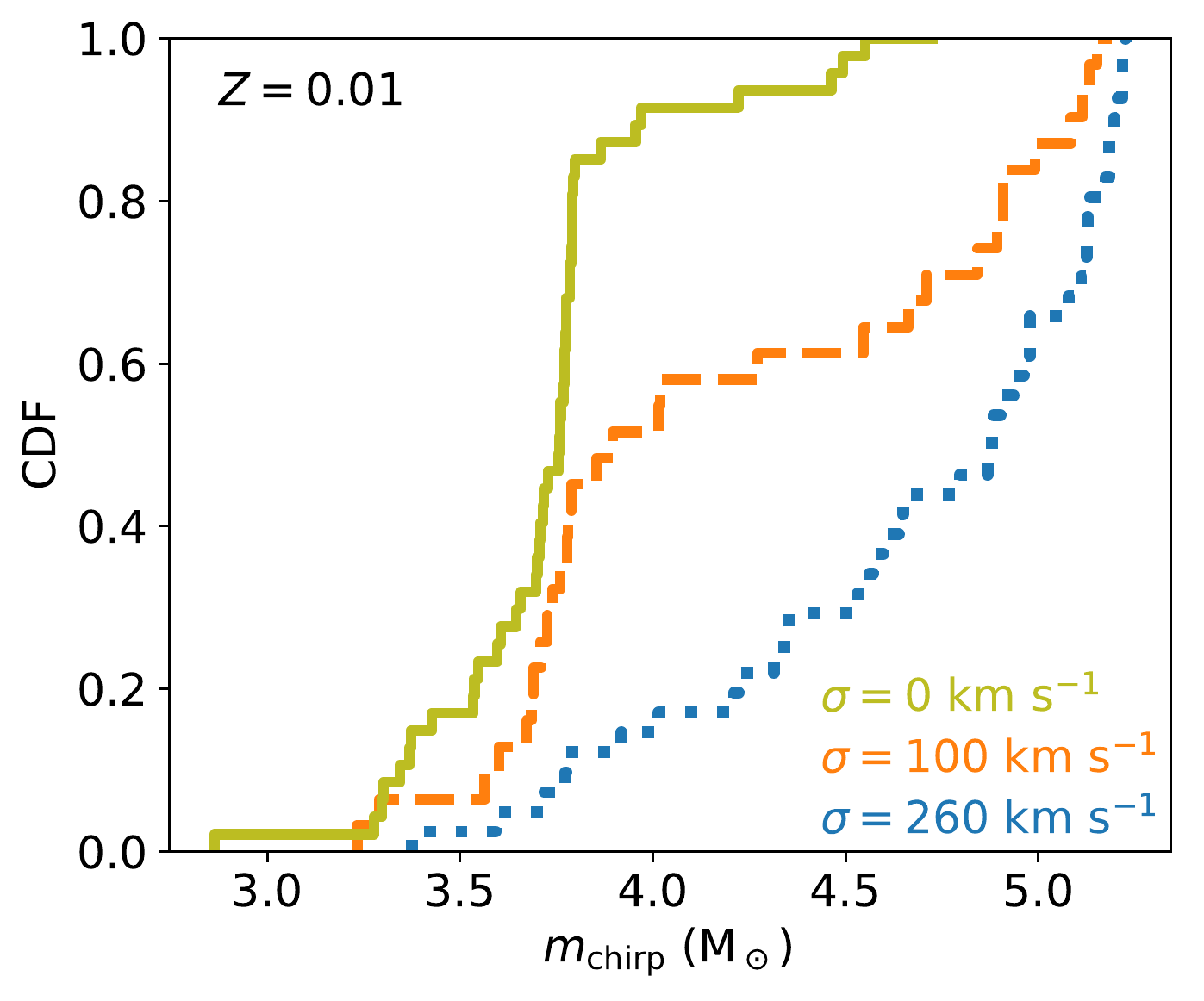}
\caption{Cumulative distribution function of the chirp mass of BH-NS binaries in triples that lead to a merger, for different values of $Z$ (top) and $\sigma$ (bottom).}
\label{fig:mchirp}
\end{figure}

Figure~\ref{fig:mchirp} shows the cumulative distribution function of the chirp mass,
\begin{equation}
m_\mathrm{chirp}=\frac{\left(\mbh\mns\right)^{3/5}}{\left(\mbh+\mns\right)^{1/5}}
\end{equation}
of BH-NS binaries in triples that lead to a merger, for different values of $Z$ (top) and $\sigma$ (bottom). As before, lower progenitor metallicities predict higher values of $m_{\rm chirp}$. For $Z\gtrsim 0.005$, the distribution of chirp masses is not significantly affected by $Z$, and most of the mergers have $m_{\rm chirp}\lesssim 5\msun$, while $m_{\rm chirp}$ is in the range $\sim 5\msun-7\msun$ and $\sim 6\msun-8\msun$ for $Z=0.001$ and $Z=0.0001$, respectively. Higher natal kicks prefer higher values of $m_{\rm chirp}$.

\subsection{Eccentricity}

\begin{figure} 
\includegraphics[scale=0.55]{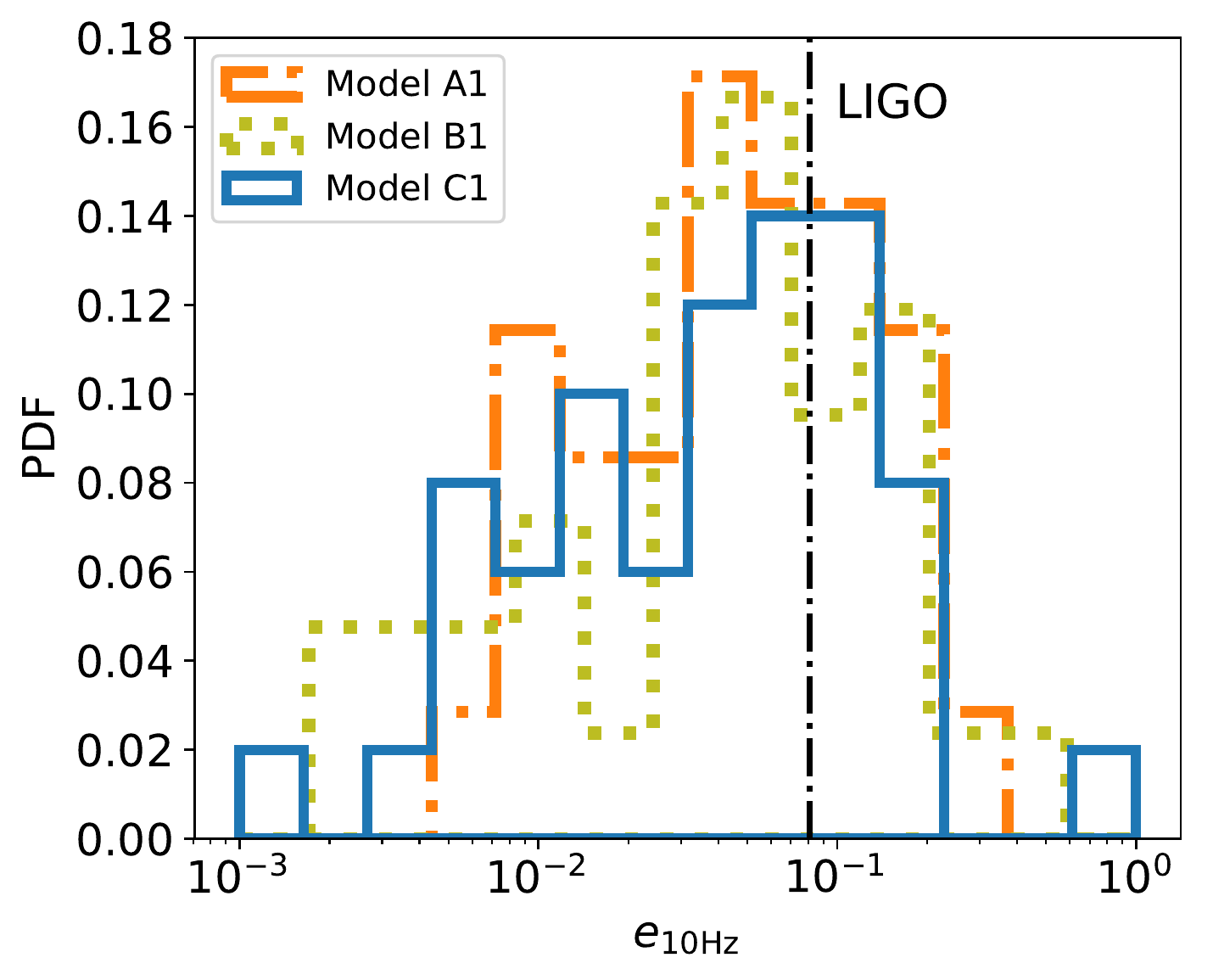}
\caption{Distribution of eccentricities at the moment the BH-NS binaries enter the LIGO frequency band ($10$ Hz). The vertical line reports the minimum eccentricity ($e_{\rm 10Hz}=0.081$) where LIGO may start distinguishing eccentric sources from circular sources \citep{gond2019}.}
\label{fig:ecc}
\end{figure}

Hierarchical configurations are expected to have eccentricities when entering the LIGO band ($10$ Hz) that are larger than for binaries that merge in isolation \citep[see e.g.][]{antchrod2016,fragrish2018,frbr2019}. For the BH binaries that merge in our simulations, we compute a proxy for the GW frequency, i.e. the frequency corresponding to the harmonic that gives the maximum GW emission \citep{wen03},
\begin{equation}
f_{\rm GW}=\frac{\sqrt{G(\mbh+\mns)}}{\pi}\frac{(1+\ein)^{1.1954}}{[\ain(1-e_{\rm in}^2)]^{1.5}}\ .
\end{equation}
Figure~\ref{fig:ecc} illustrates the distribution (PDF) of eccentricities at the moment the BH binaries enter the LIGO frequency band. We also plot the minimum $e_{\rm 10Hz}=0.081$ where LIGO could start distinguishing eccentric sources from circular sources \citep{gond2019}. A large fraction of systems that merge have a significant eccentricity in the LIGO band, compared to binaries that merge in isolation. We note that a similar signature could be found in BH binaries that merge near supermassive black holes \citep{fragrish2018,flp2019}, in the BH-BH binaries that merge in star clusters \citep{2014sams,sam2018,2019sams}, and in isolated hierarchical triples \citep{ant17} and quadruples \citep{fragk2019}.

\subsection{Effective spin}

We describe the BH and NS spins with their respective spin vectors ${\bf{S_{\rm BH}}}=S_{\rm BH}{\bf{\hat{S}_{\rm BH}}}$ and ${\bf{S_{\rm NS}}}=S_{\rm NS}{\bf{\hat{S}_{\rm NS}}}$, where $S_{\rm BH}=(Gm_{\rm BH}^2/c)\chi_{\rm BH}$ and where $S_{\rm NS}=(Gm_{\rm NS}^2/c)\chi_{\rm BH}$. In the previous equations, $0\le \chi_{\rm BH}\le 1$ and $0\le \chi_{\rm NS}\le 1$ are the Kerr parameter of the BH and NS, respectively. Rather then the single spins of the two merging objects, GW measurements are sensible to a combination of them,
\begin{equation}
\chieff=\frac{\mbh\chibh\cos\thetabh+\mns\chins\cos\thetans}{\mbh+\mns}\ ,
\end{equation}
where $\cos\thetabh=({\bf{\hat{S}_{\rm BH}}\cdot \bf{J}})/J$ and $\cos\thetans=({\bf{\hat{S}_{\rm NS}}\cdot \bf{J}})/J$, where $\bf{J}$ is the BH-NS orbital angular momentum.

The spins of the BHs and NSs in the inner binary can undergo a relativistic precession around $\bf{J}$, as a consequence of the KL cycles. In our simulations, we model the spin-orbit coupling as in \citet{liulai2019}. The De Sitter geodetic precession of the spins ${\bf{S_{\rm BH}}}$ and ${\bf{S_{\rm NS}}}$ of the BHs in the inner binary around the inner binary angular momentum $\bf{J}$ is given by,
\begin{equation}
\frac{d\bf{S_{\rm BH}}}{dt}={\bf{\Omega_{\rm BH}}}\times {\bf{S_{\rm BH}}}=\left[\frac{2G\mu}{c^2 r^3}\left(1+\frac{3\mns}{4\mbh}\right){\bf{r}}\times {\bf{v}}\right] \times {\bf{S_{\rm BH}}}
\end{equation}
\begin{equation}
\frac{d\bf{S_{\rm NS}}}{dt}={\bf{\Omega_{\rm NS}}}\times {\bf{S_{\rm NS}}}=\left[\frac{2G\mu}{c^2 r^3}\left(1+\frac{3\mbh}{4\mns}\right){\bf{r}}\times {\bf{v}}\right] \times {\bf{S_{\rm NS}}} \ ,
\end{equation}
where $\mu$ is the inner binary reduced mass, ${\bf{r}}={\bf{r_{\rm BH}}}-{\bf{r_{\rm NS}}}$ and ${\bf{v}}={\bf{v_{\rm BH}}}-{\bf{v_{\rm NS}}}$\footnote{We neglect the backreaction of $\bf{S_{\rm BH}}$ and $\bf{S_{\rm NS}}$ on $\bf{J}$ and the spin-spin precessional terms \citep{antonini2018,liulai2019}.}.

\begin{table}
\caption{Spin models: name, Kerr parameter of the BH ($\chibh$), Kerr parameter of the NS ($\chins$), initial misalignment of the BH spin ($\cos\theta_{\rm BH}^{\rm ini}$) and NS spin($\cos\theta_{\rm NS}^{\rm ini}$) with respect to $\bf{J}$.}
\centering
\begin{tabular}{lccc}
\hline
Name & $\chibh$ & $\chins$ & Initial misalignment\\
\hline\hline
S1 & uniform              & uniform & $0^\circ \le\cos\theta_{\rm BH,NS}^{\rm ini}\le 20^\circ$  \\
S2 & Eq.~\ref{eqn:bhspin} & uniform & $0^\circ \le\cos\theta_{\rm BH,NS}^{\rm ini}\le 20^\circ$ \\
T1 & uniform              & uniform & aligned to $\bf{J}$ \\
T2 & uniform              & uniform & isotropic \\
U1 & $0.2$                & $0.2$   & $0^\circ \le\cos\theta_{\rm BH,NS}^{\rm ini}\le 20^\circ$ \\
U2 & $0.5$                & $0.5$   & $0^\circ \le\cos\theta_{\rm BH,NS}^{\rm ini}\le 20^\circ$ \\
U3 & $0.8$                & $0.8$   & $0^\circ \le\cos\theta_{\rm BH,NS}^{\rm ini}\le 20^\circ$ \\
\hline
\end{tabular}
\label{tab:spins}
\end{table}

\begin{figure*} 
\centering
\includegraphics[scale=0.55]{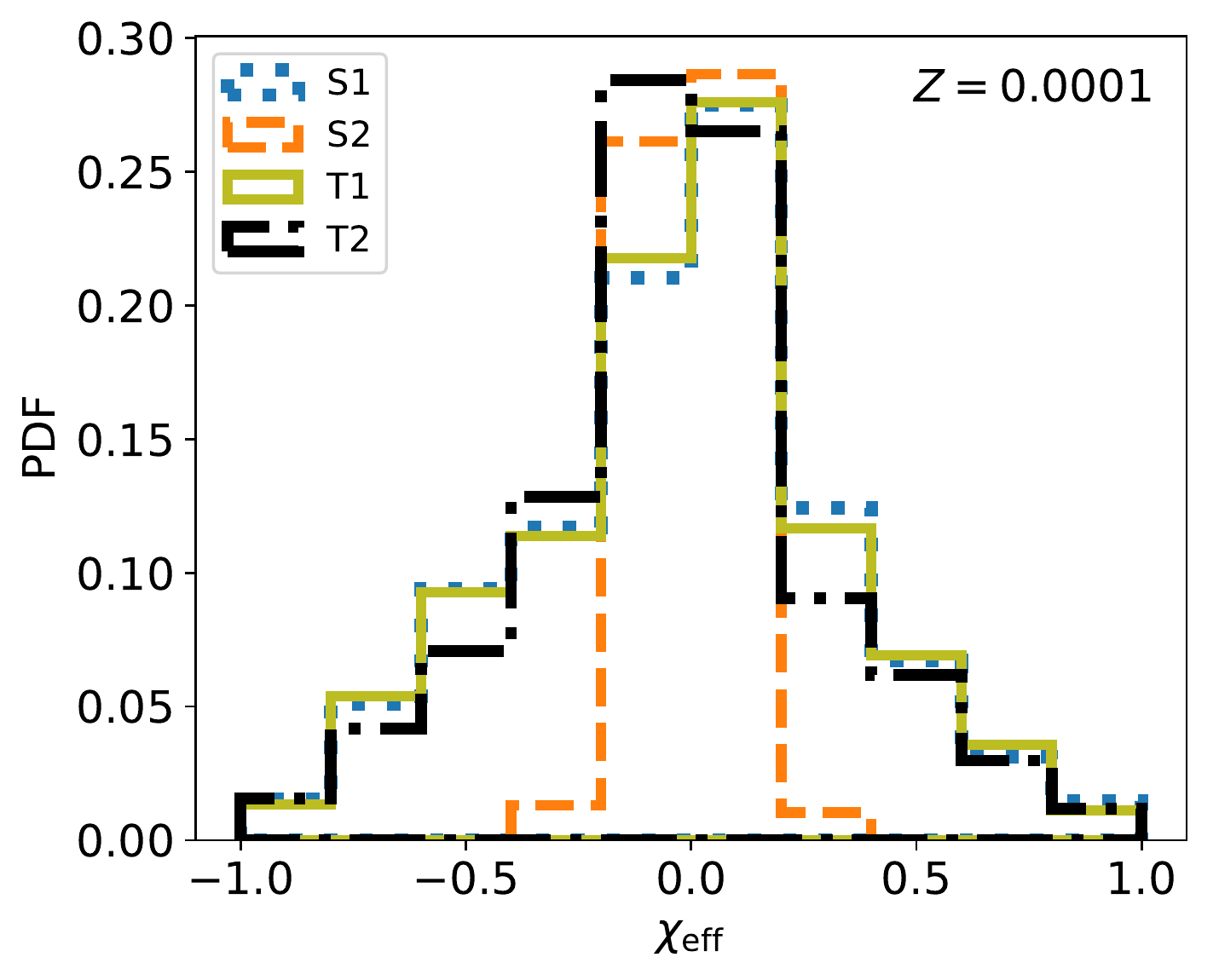}
\hspace{0.5cm}
\includegraphics[scale=0.55]{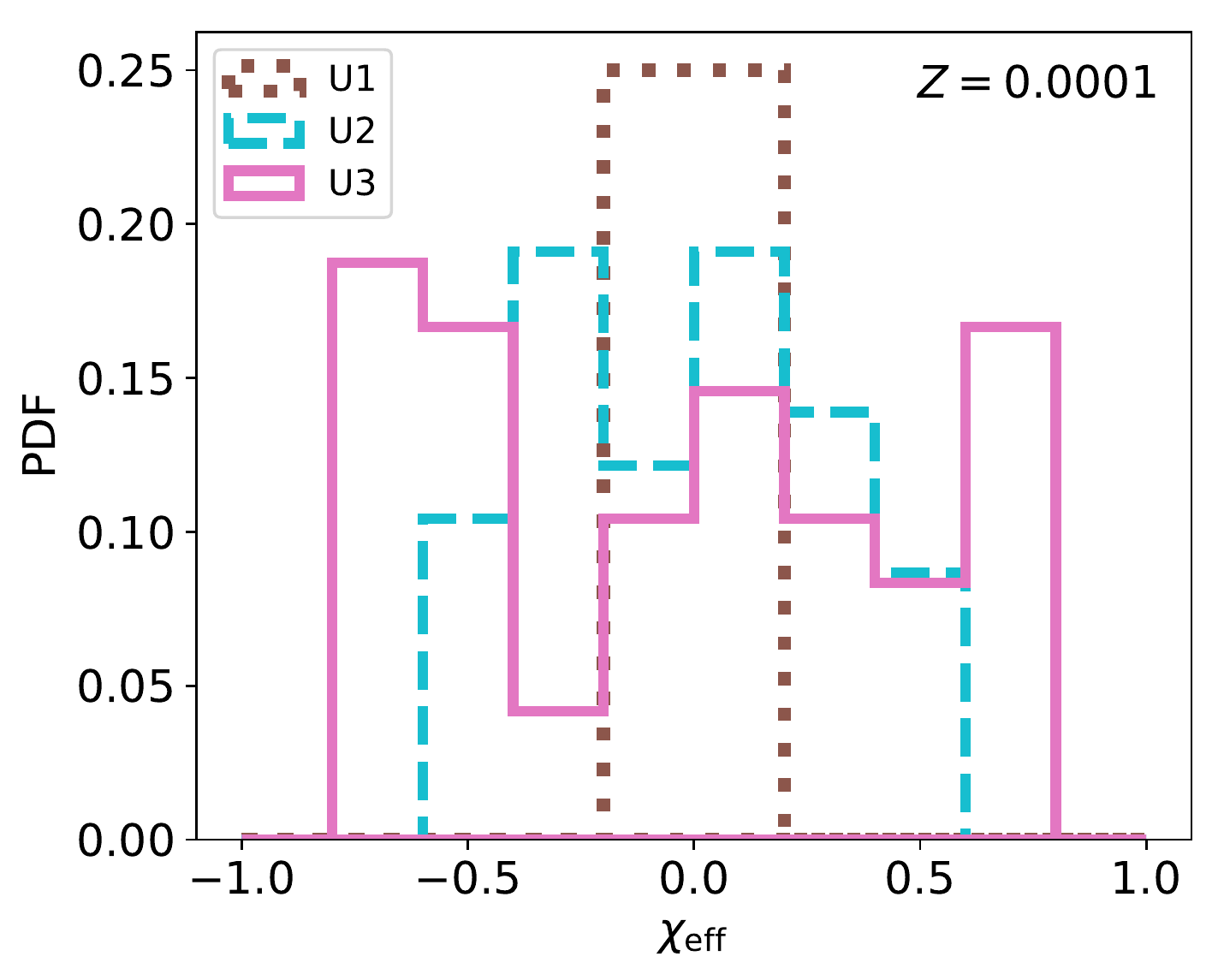}
\includegraphics[scale=0.55]{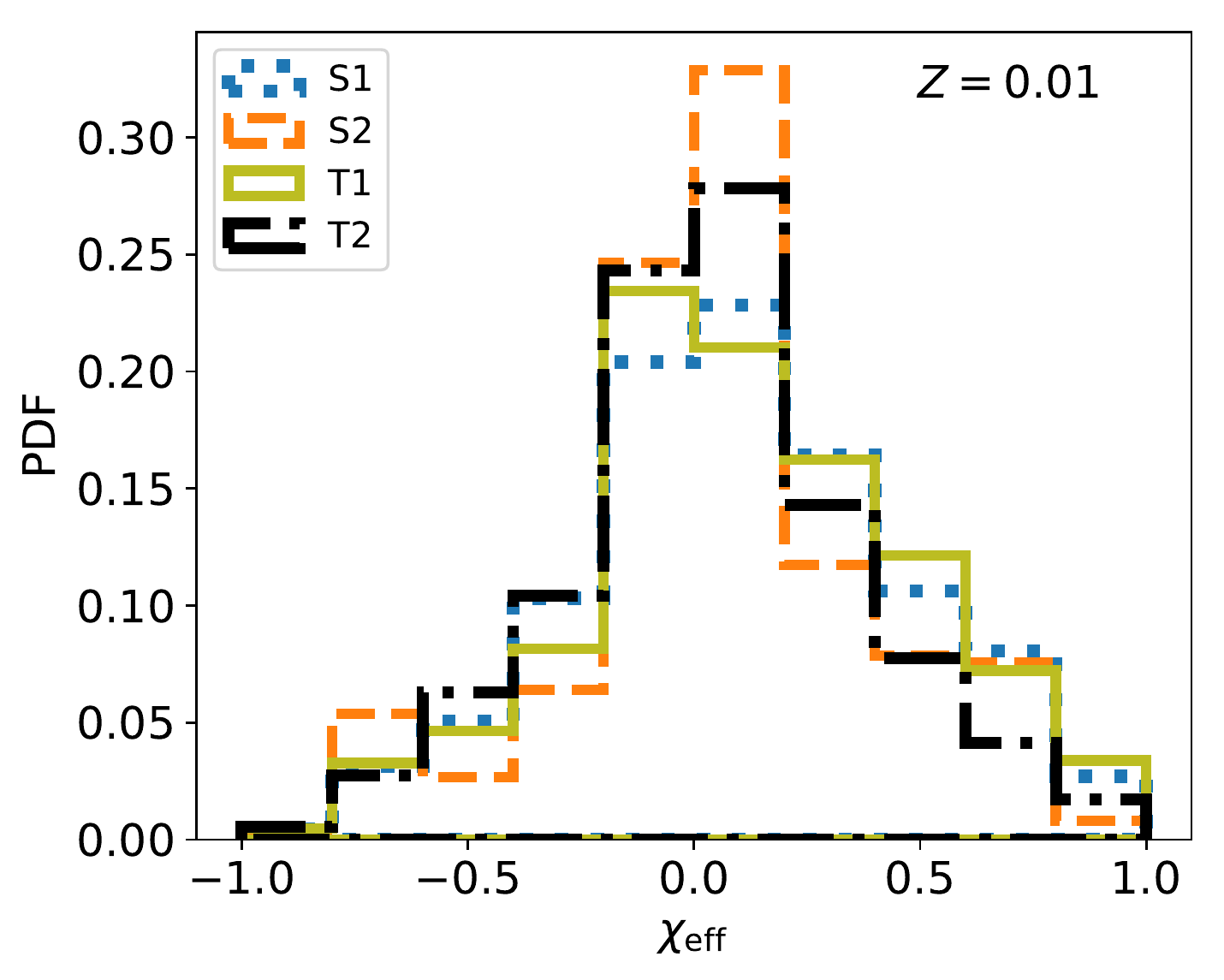}
\hspace{0.5cm}
\includegraphics[scale=0.55]{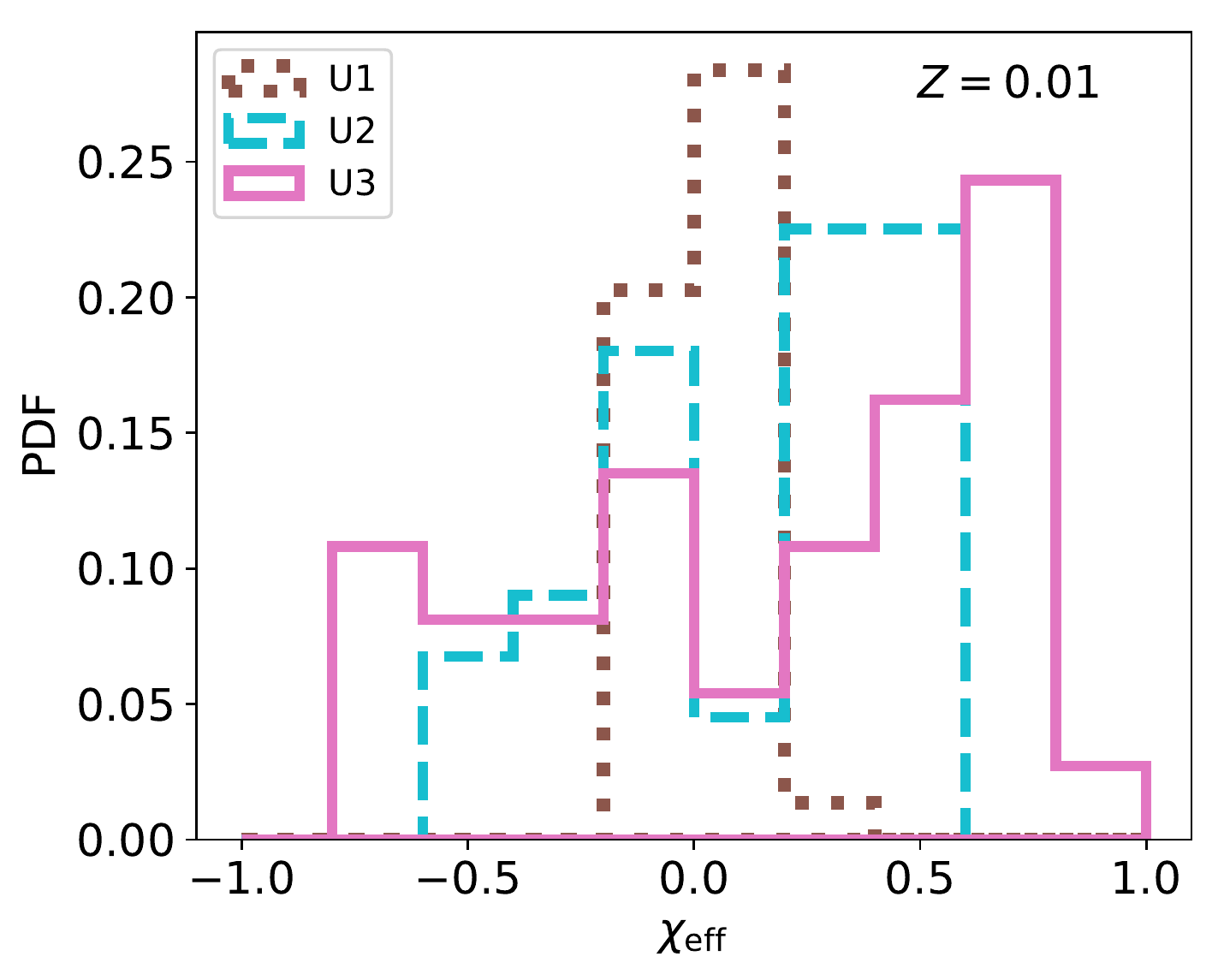}
\includegraphics[scale=0.55]{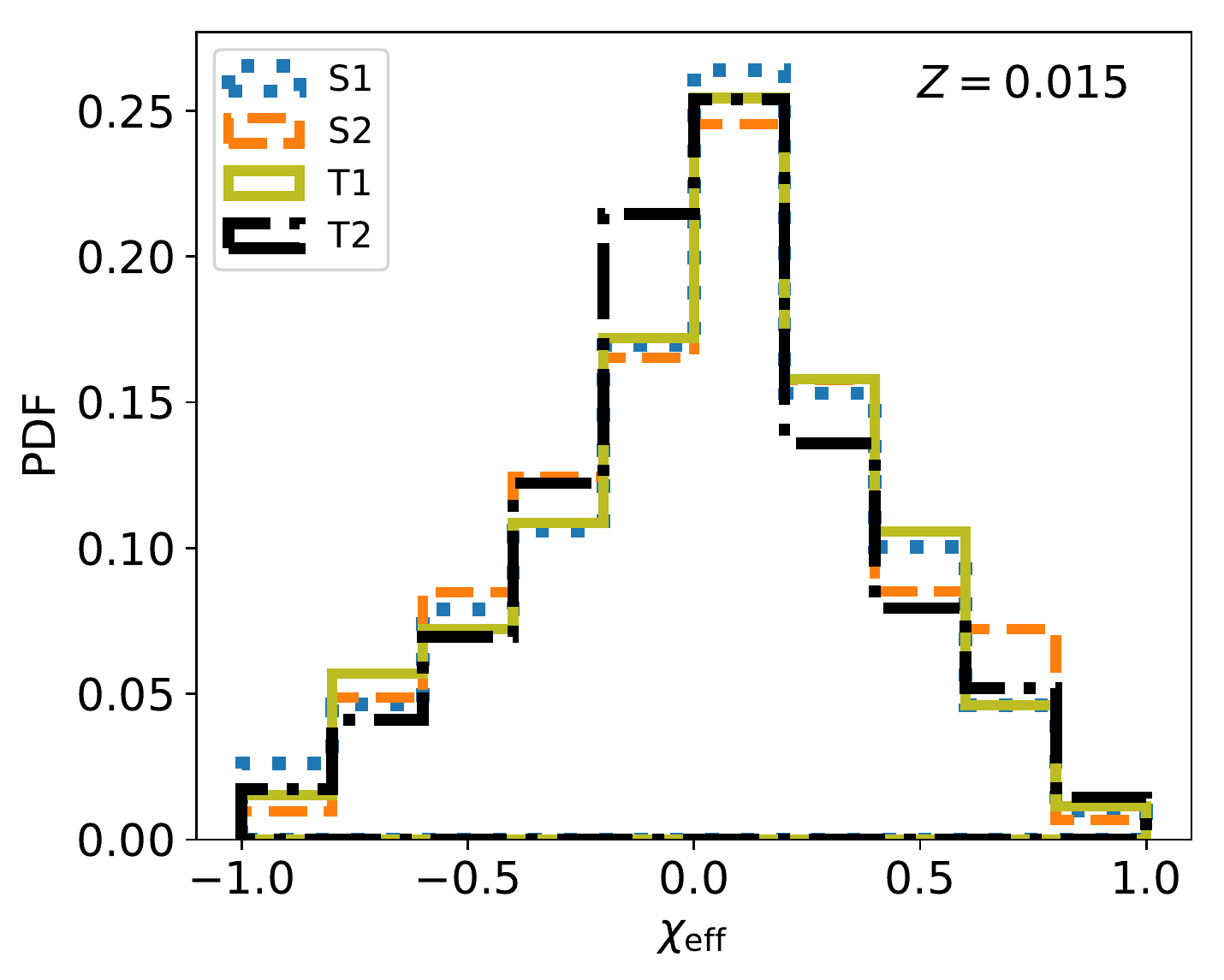}
\hspace{0.5cm}
\includegraphics[scale=0.55]{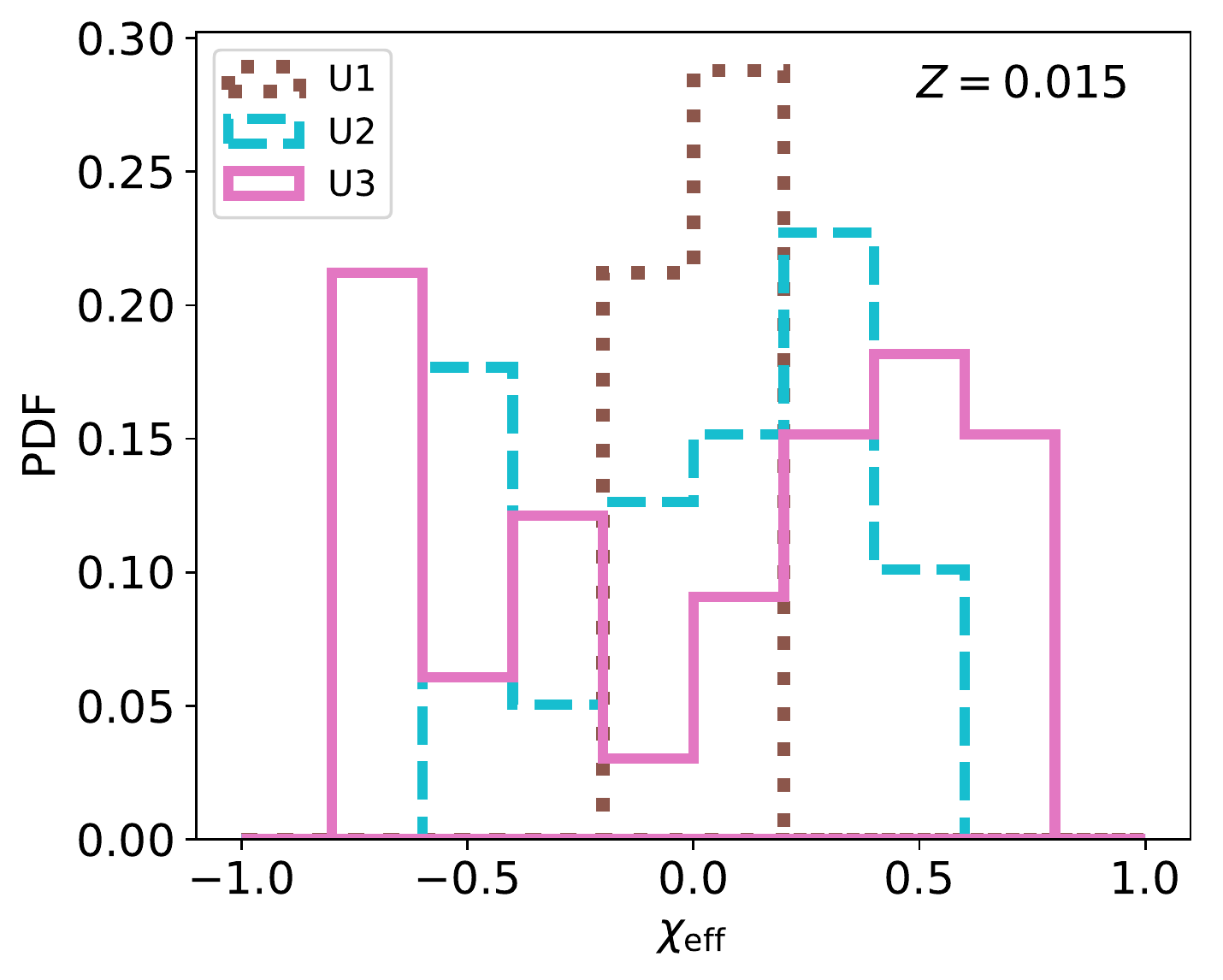}
\caption{Effective spin distributions of BH-NS binaries in triples that lead to merger for different values of $Z$ (top: Model B1; centre: Model A1; bottom: Model B4) and all the spin models under consideration (see Table~\ref{tab:spins}).}
\label{fig:spin1}
\end{figure*}

\begin{figure*} 
\centering
\includegraphics[scale=0.55]{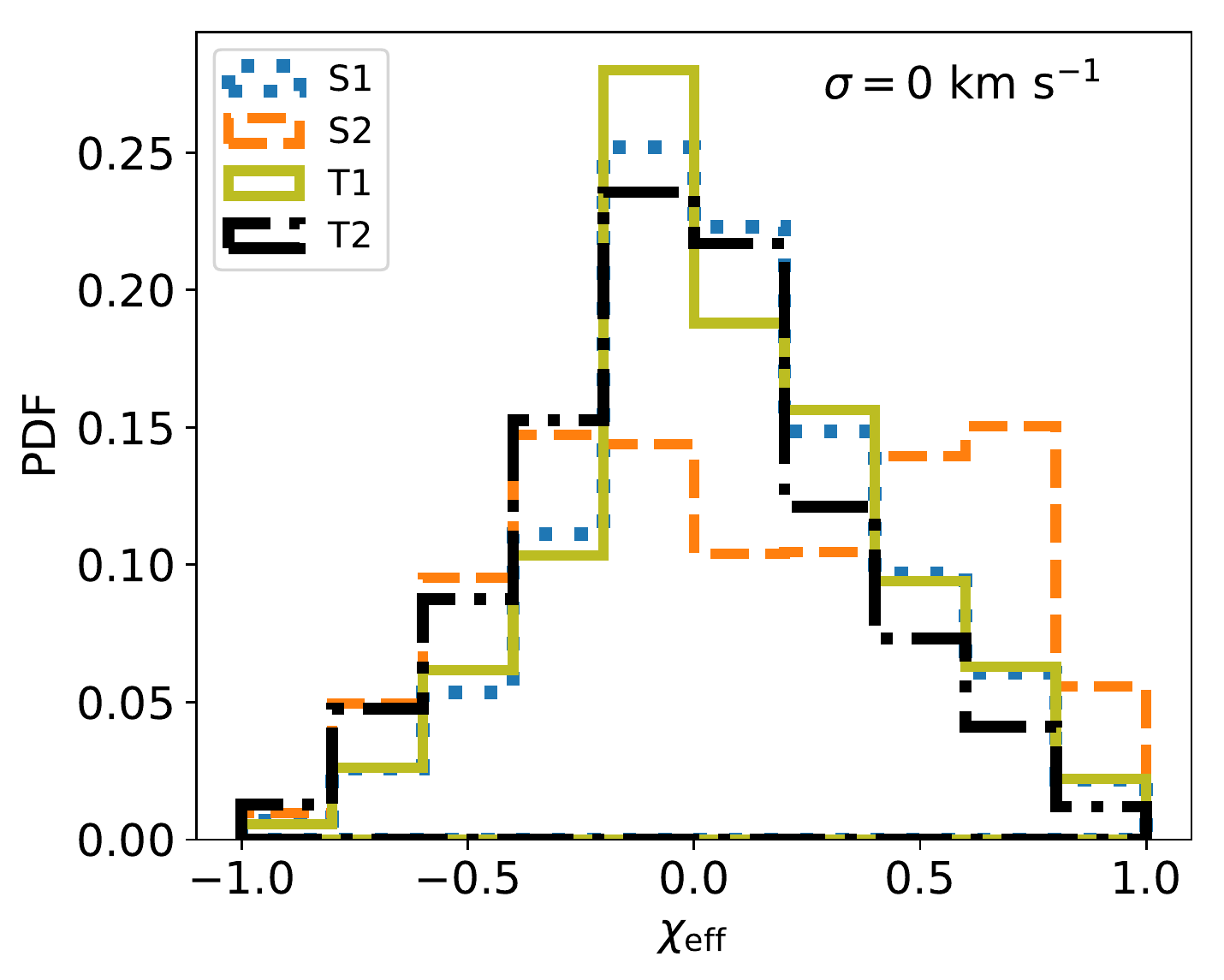}
\hspace{0.5cm}
\includegraphics[scale=0.55]{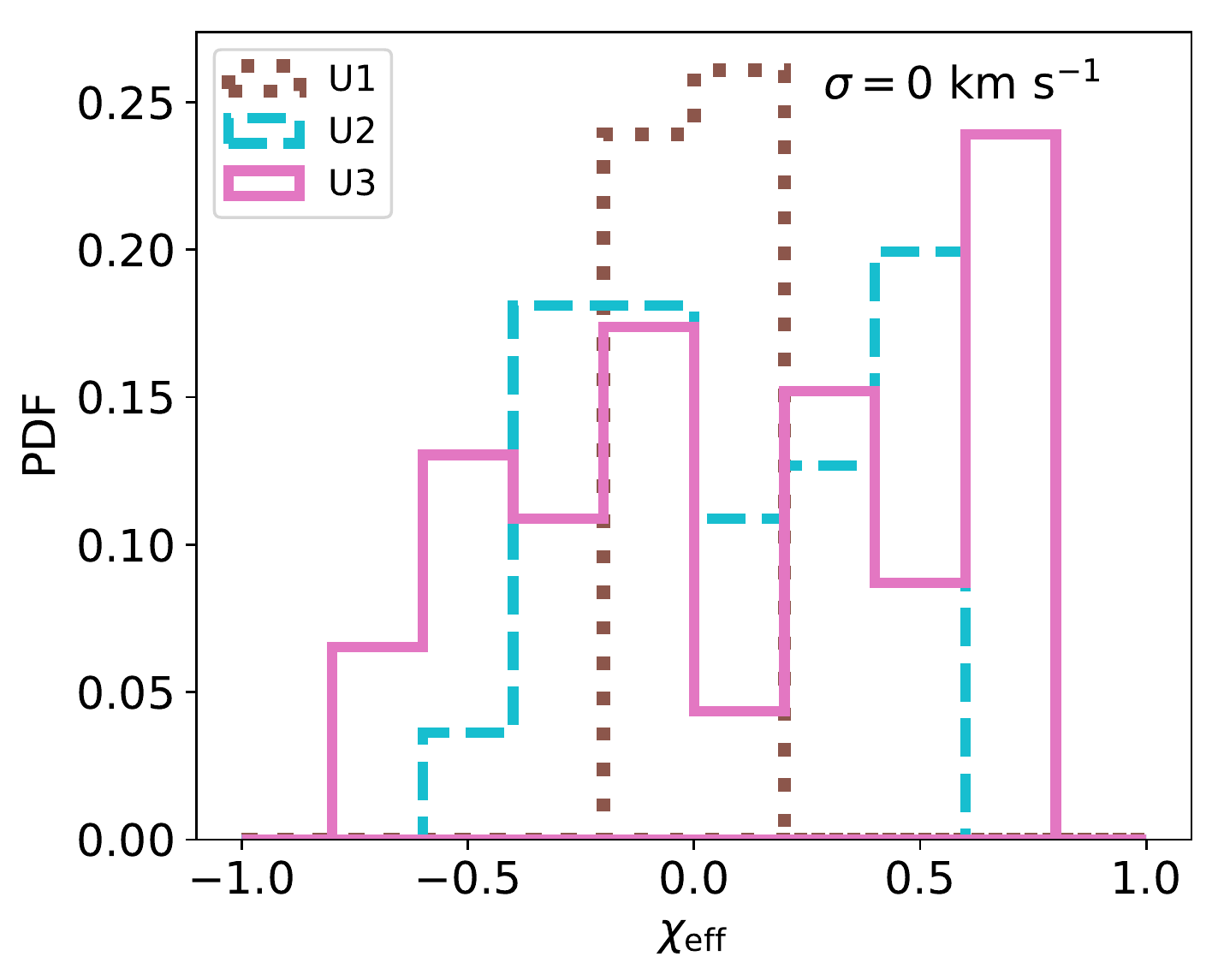}
\includegraphics[scale=0.55]{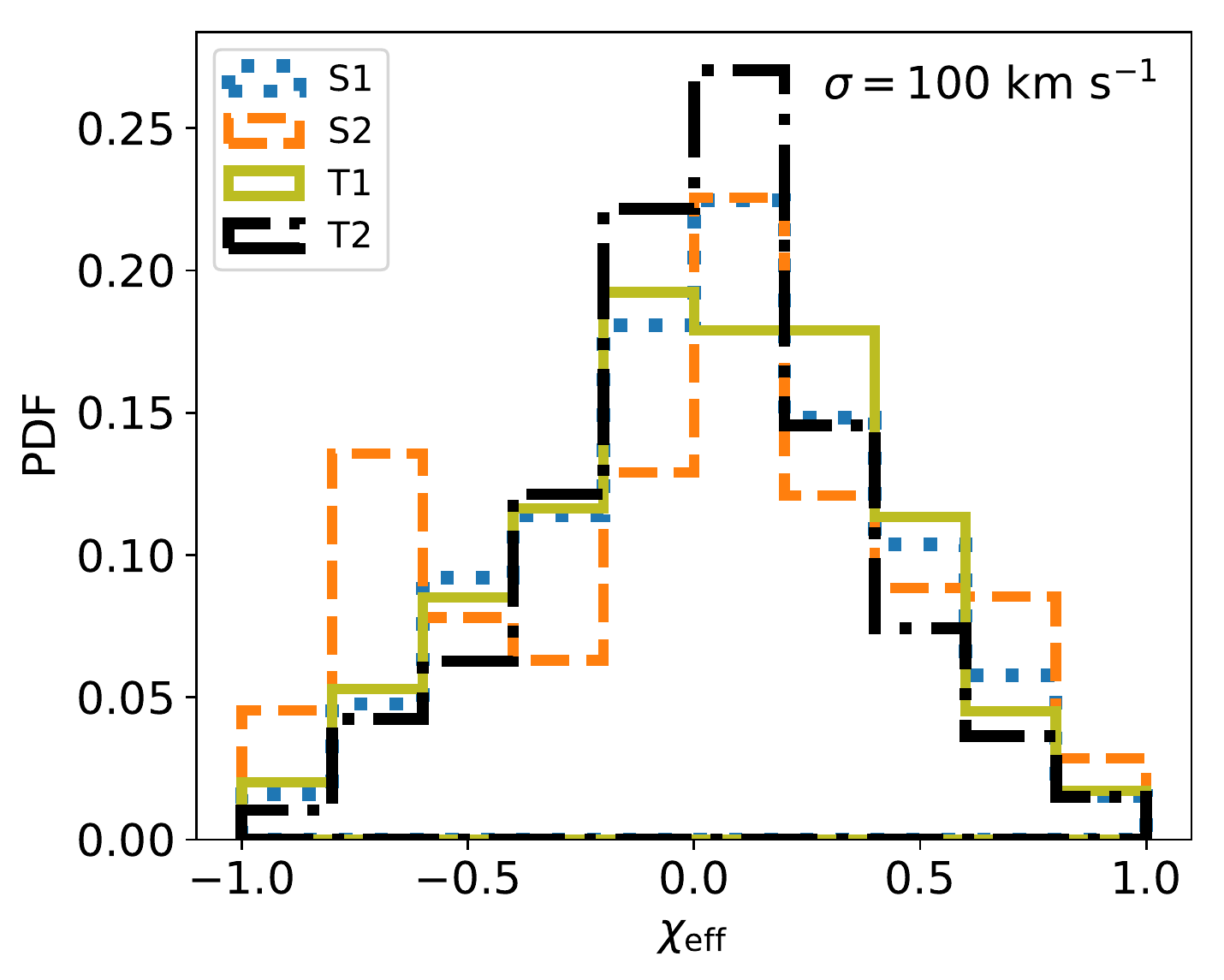}
\hspace{0.5cm}
\includegraphics[scale=0.55]{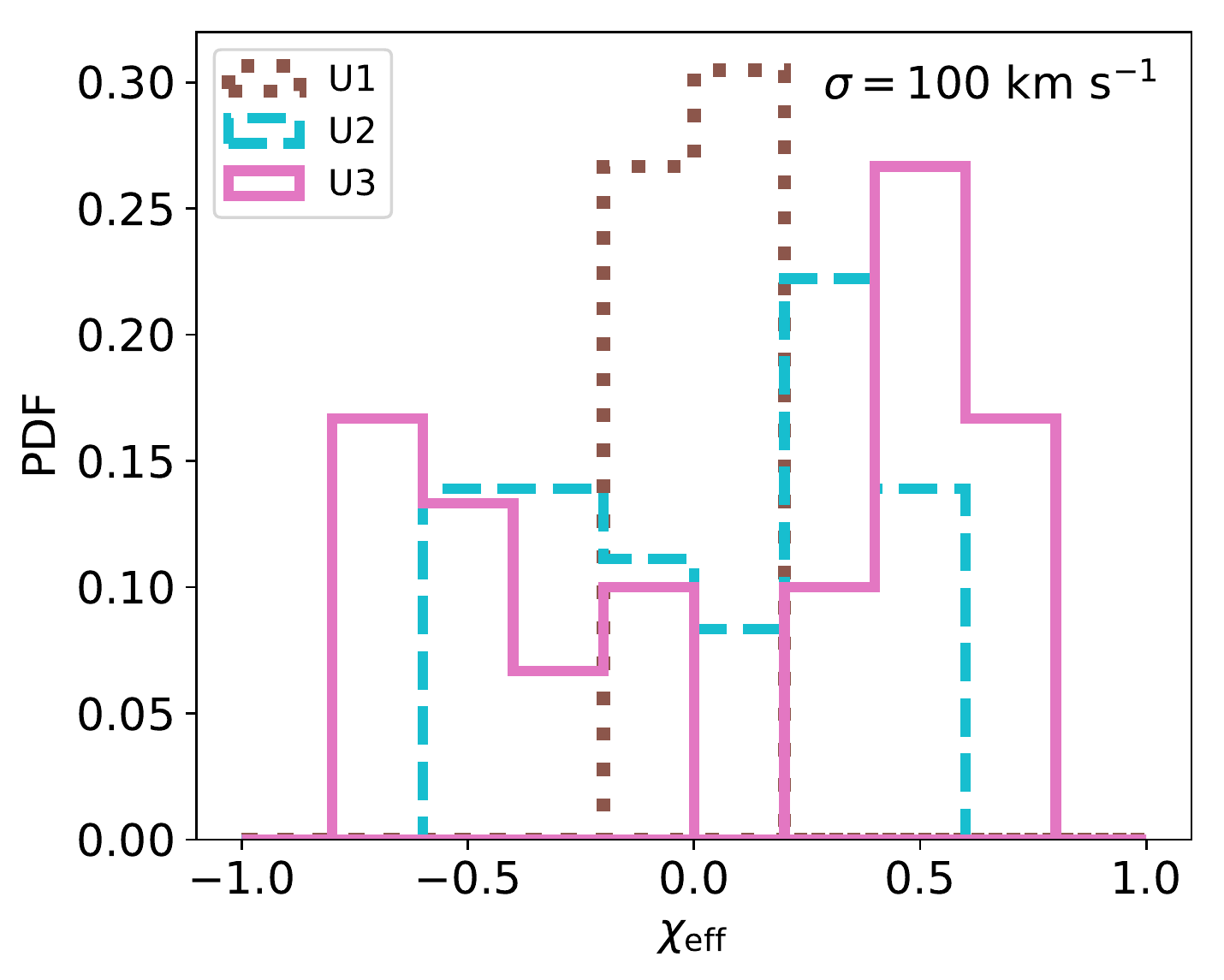}
\includegraphics[scale=0.55]{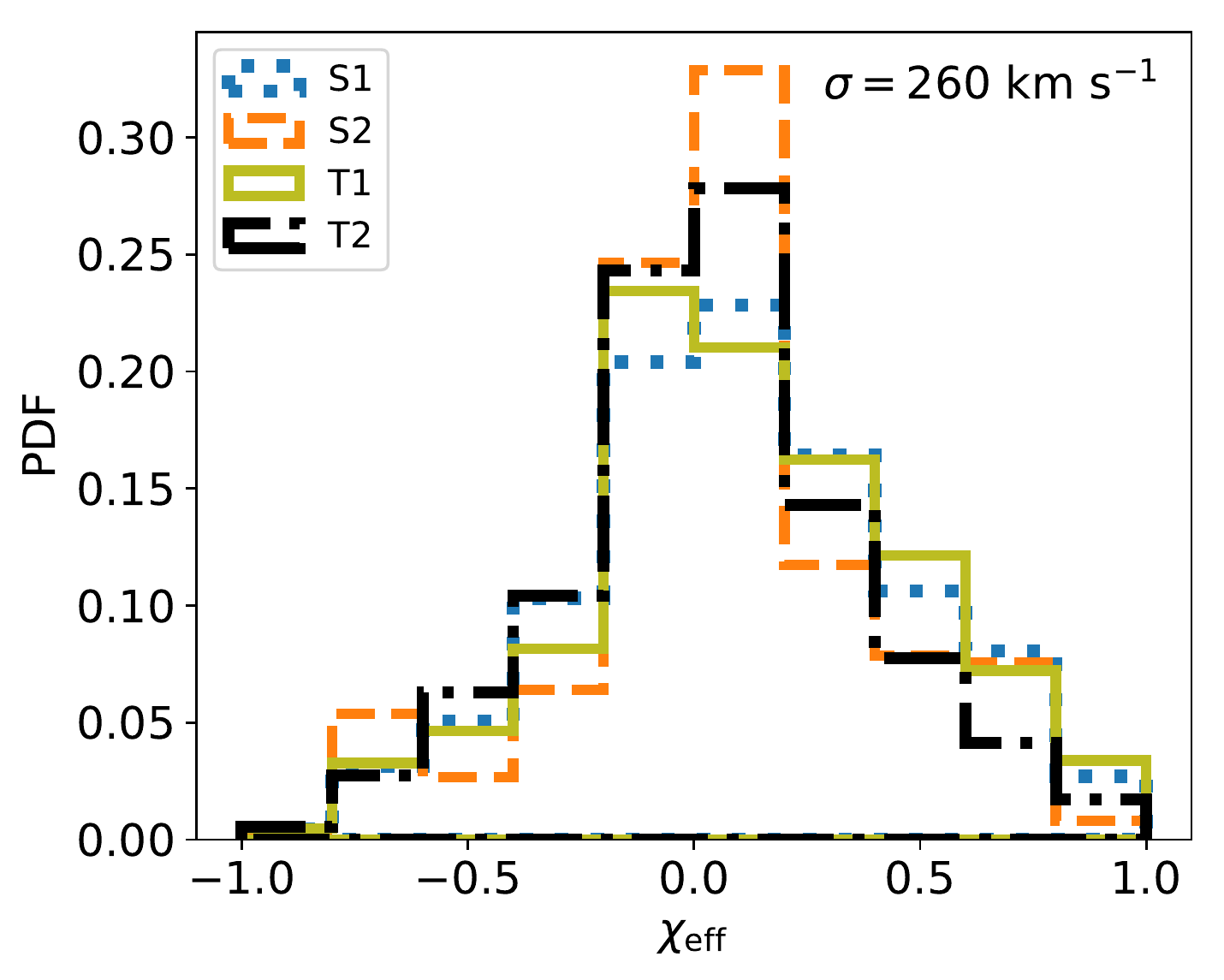}
\hspace{0.5cm}
\includegraphics[scale=0.55]{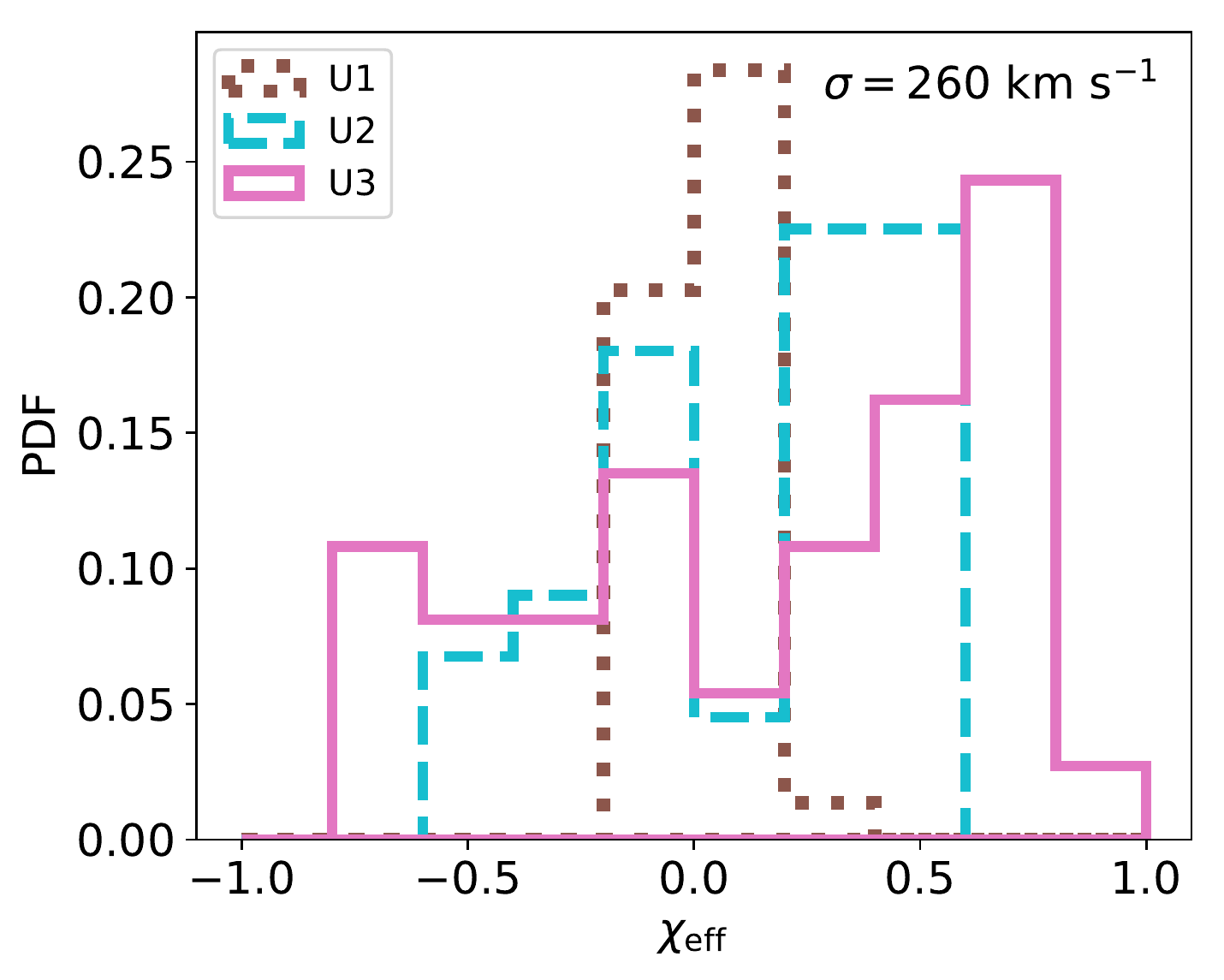}
\caption{Effective spin distributions of BH-NS binaries in triples that lead to merger for different values of $\sigma$ (top: Model A3; centre: Model A2; bottom: Model A1) and all the spin models under consideration (see Table~\ref{tab:spins}).}
\label{fig:spin2}
\end{figure*}

\begin{figure*} 
\centering
\includegraphics[scale=0.55]{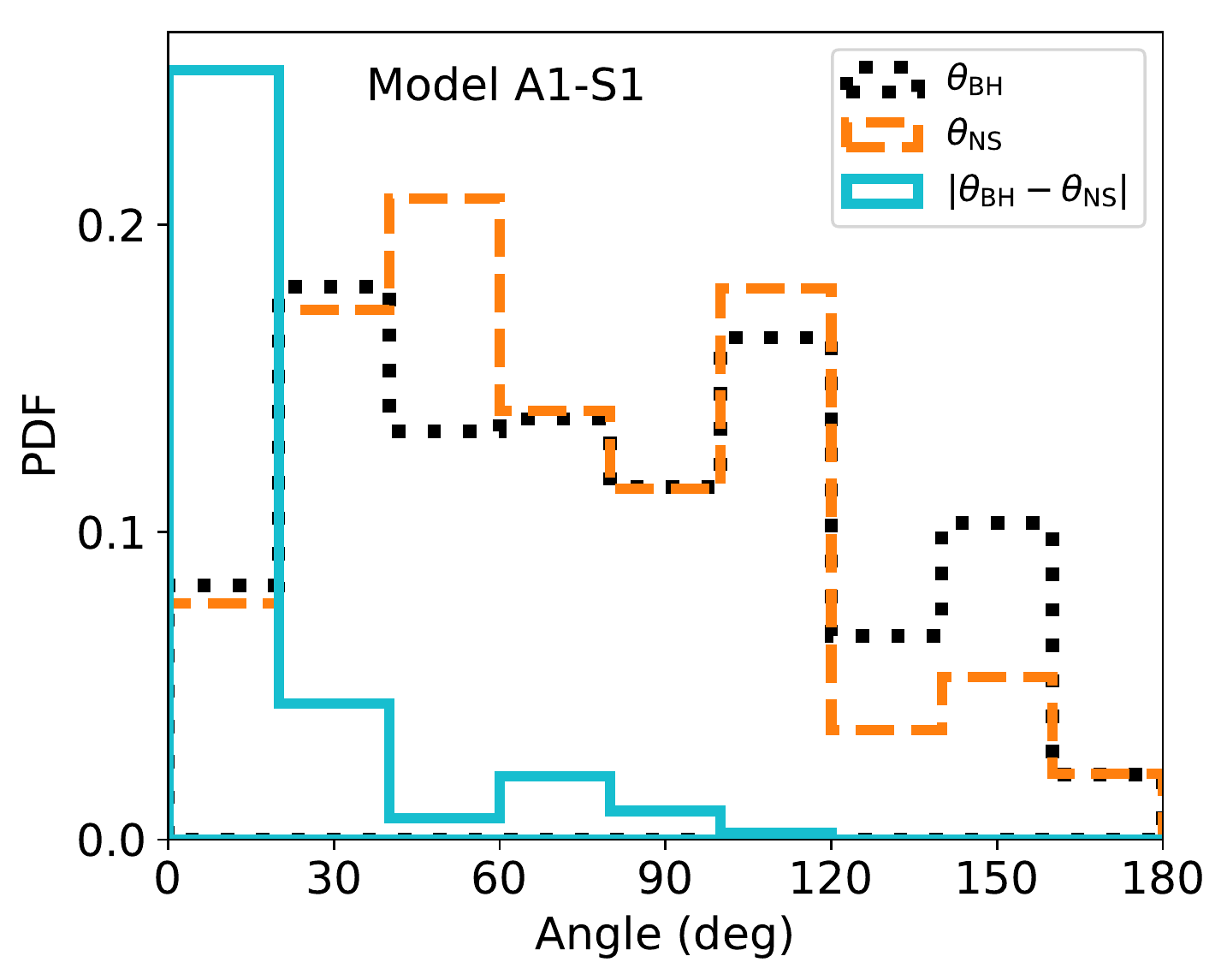}
\hspace{0.5cm}
\includegraphics[scale=0.55]{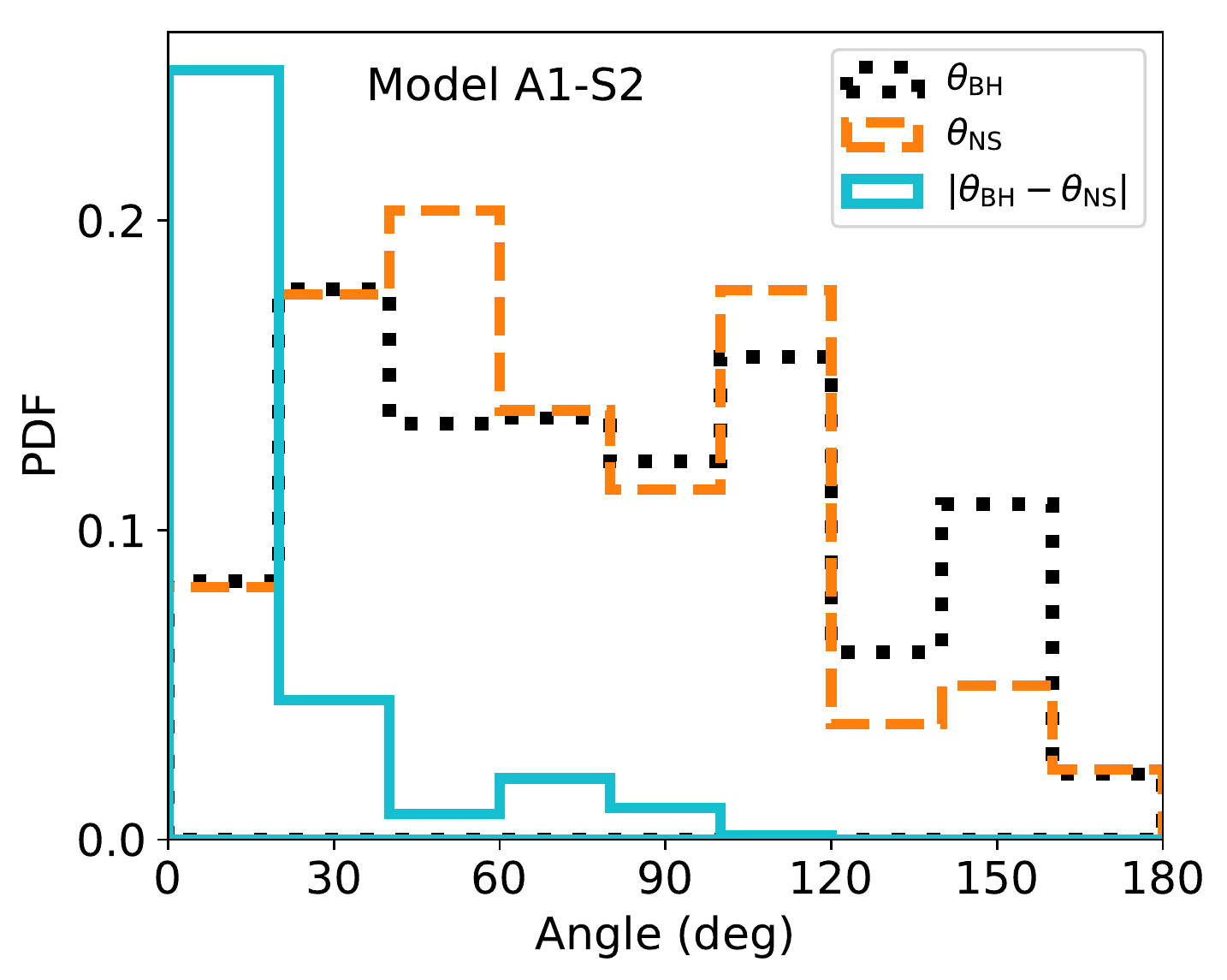}
\includegraphics[scale=0.55]{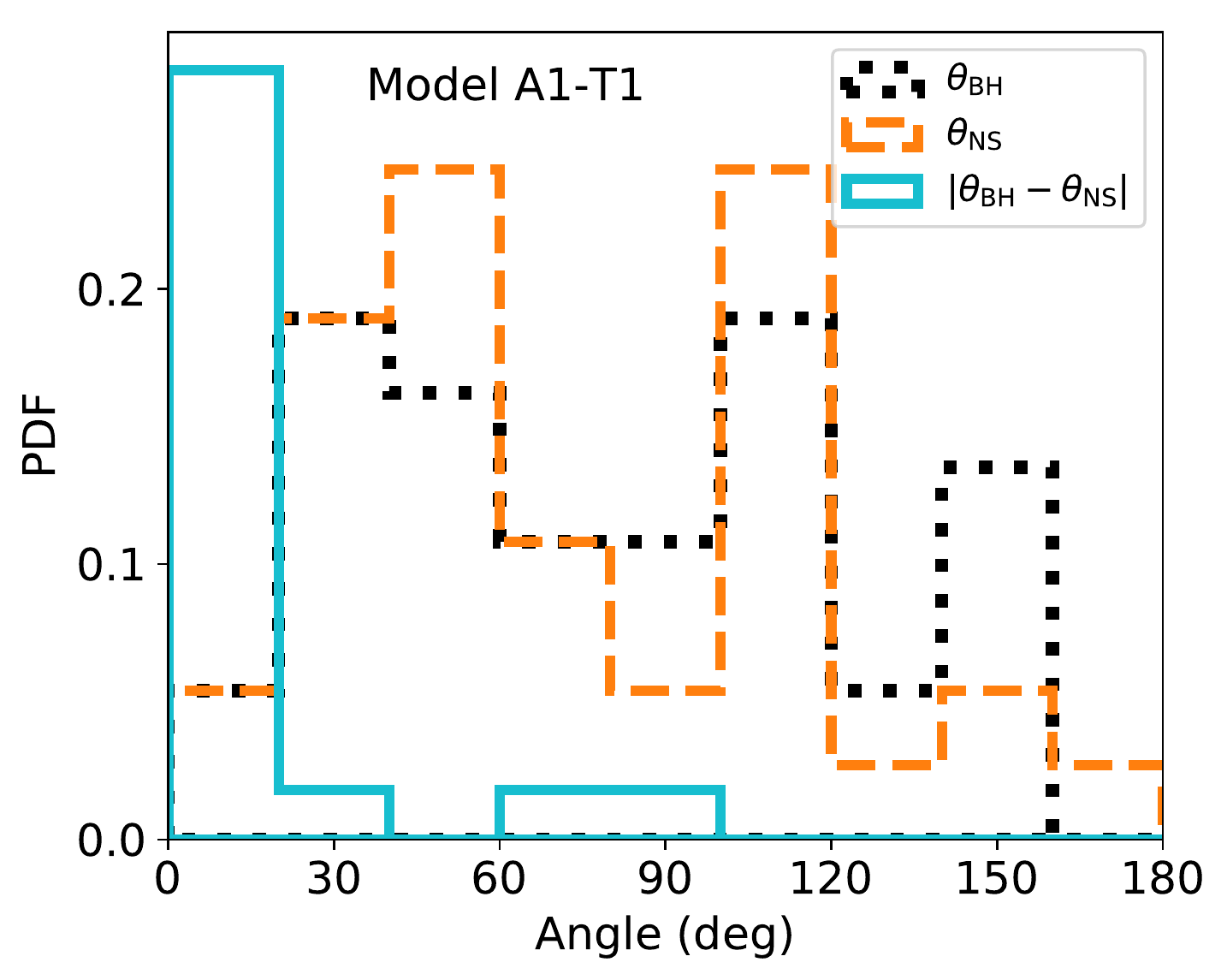}
\hspace{0.5cm}
\includegraphics[scale=0.55]{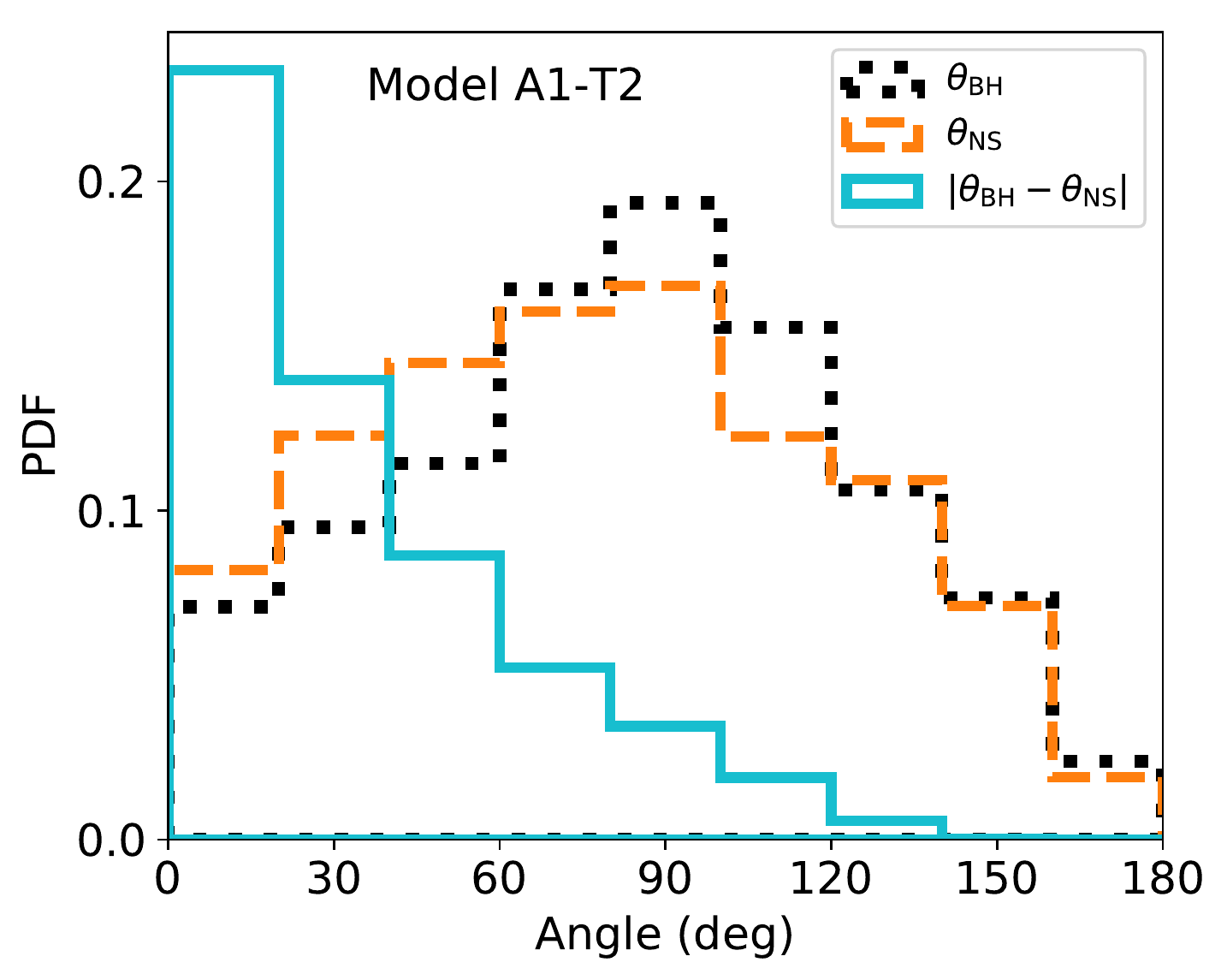}
\caption{Distributions of the absolute misalignment $\thetabh$ between ${\bf{S_{\rm BH}}}$ and $\bf{J}$, $\thetans$ between ${\bf{S_{\rm NS}}}$ and $\bf{J}$, and the relative misalignment between ${\bf{S_{\rm BH}}}$ and ${\bf{S_{\rm NS}}}$ for Model A1 and the first four spin models of Table~\ref{tab:spins}.}
\label{fig:spin3}
\end{figure*}

We take seven different models for the spins, which differ for the initial magnitude of the Kerr parameters and/or the initial orientation of the spins (see Table~\ref{tab:spins})
\begin{itemize}
\item Model S1: $\chibh$ and $\chins$ are drawn independently from an uniform distribution and the initial misalignments of the BH spin ($\cos\theta_{\rm BH}^{\rm ini}$) and NS spin($\cos\theta_{\rm NS}^{\rm ini}$) with respect to $\bf{J}$ are drawn uniformly in the range $0^\circ$--$20^\circ$;
\item Model S2: the dimensionless BH spin is set by the BH mass \citep{belc2017},
\begin{equation}
\chi=\frac{p_1-p_2}{2}\tanh\left(p_3-\frac{\mbh}{\msun}\right)+\frac{p_1+p_2}{2}\ ,
\label{eqn:bhspin}
\end{equation}
where $p_1=0.86\pm 0.06$, $p_2=0.13\pm 0.13$, and $p_3=29.5\pm 8.5$. Spins are then generated by drawing random samples uniformly in the region in between the two curves given by the upper and lower limits of the parameters \citep{gerosa2018}. The dimensionless NS spin is drawn from an uniform distribution, while $0^\circ \le\cos\theta_{\rm BH,NS}^{\rm ini}\le 20^\circ$ uniformly;
\item Model T1: $\chibh$ and $\chins$ are drawn independently from an uniform distribution and the BH and NS spins are initially aligned to the BH-NS angular momentum;
\item Model T2: $\chibh$ and $\chins$ are drawn independently from an uniform distribution and the initial spin-orbit misalignents of BH and NS are drawn from an isotropic distribution;
\item Models U1, U2, U3: the BH and NS Kerr parameters are fixed to $0.2$--$0.5$--$0.8$, respectively, and $0^\circ \le\cos\theta_{\rm BH,NS}^{\rm ini}\le 20^\circ$ uniformly.
\end{itemize}

In Figure~\ref{fig:spin1}, we show the effective spin distributions of BH-NS binaries in triples that lead to merger for different values of $Z$ (top: Model B1; centre: Model A1; bottom: Model B4) and all the spin models in Table~\ref{tab:spins}. In Models S1-S2-T1-T2 (left panel), the $\chieff$ distribution is not affected by the initial choice of the BH and NS spins, for $Z=0.01$ and $Z=0.015$. The distributions have a peak at $\chieff\sim 0$ and broad tails up to $\pm 1$. We find a similar trend for all the spin models also for $Z=0.0001$, except for the Model S2. In these model, we sample $\chibh$ according to Eq.~\ref{eqn:bhspin}, where more massive BHs have on average smaller initial spins. Moreover, for $Z=0.0001$, the final mass of the BH (see Figure~\ref{fig:mass}) may be considerably higher than the mass of the NS. As a result, $\chieff$ could be mainly determined by the BH contribution, and only slightly affected by the NS, thus result in a distribution peaked at $\chieff\sim 0$ and negligible tails.

In the spin models where we fix the BH and NS Kerr parameters (right panel), the final distributions of $\chieff$ do not depend substantially on the progenitor metallicity and present a similar behaviour for all $Z$'s. The distributions appear more broadly distributed over the possible values of $\chieff$ for larger initial Kerr parameters. As the initial spin magnitudes are decreased, the distributions converge to zero.

Figure~\ref{fig:spin2} illustrates the distributions of $\chieff$ of BH-NS binaries in triples that lead to merger for different values of $\sigma$ (top: Model A3; centre: Model A2; bottom: Model A1) and all the spin models in Table~\ref{tab:spins}. Also here, in Models S1-S2-T1-T2 (left panel), the $\chieff$ distribution is not affected by the initial choice of the BH and NS spins. The only exception is the $\chieff$ distribution for Model S2 in the case $\sigma=0\kms$, where the distribution appears more flat. In the models where we fix the BH and NS Kerr parameters (right panel), the final distributions of $\chieff$ do not depend substantially on $\sigma$ and present a similar behaviour described above for different progenitor metallicities.

Finally, we find nearly no difference in the final distributions of the effective spin for Model C1, where we set $\amax=5000$ AU, compared to our fiducial Model A1.

We also show in Figure~\ref{fig:spin3} the distributions of the absolute misalignment $\thetabh$ between ${\bf{S_{\rm BH}}}$ and $\bf{J}$, $\thetans$ between ${\bf{S_{\rm NS}}}$ and $\bf{J}$, and the relative misalignment between ${\bf{S_{\rm BH}}}$ and ${\bf{S_{\rm NS}}}$ for Model A1 and the first four spin models of Table~\ref{tab:spins}. We find that for the spin Models S1-S2-T1, the distributions are similar to each other, while differences arise for Model T2. In particular, the distribution of $|\thetabh-\thetans|$ appears broader than for the other models. This is a result of the fact that the initial misalignment of the BH an NS spins is drawn from an isotropic distribution.

\subsection{Merger times}

\begin{figure*} 
\centering
\includegraphics[scale=0.55]{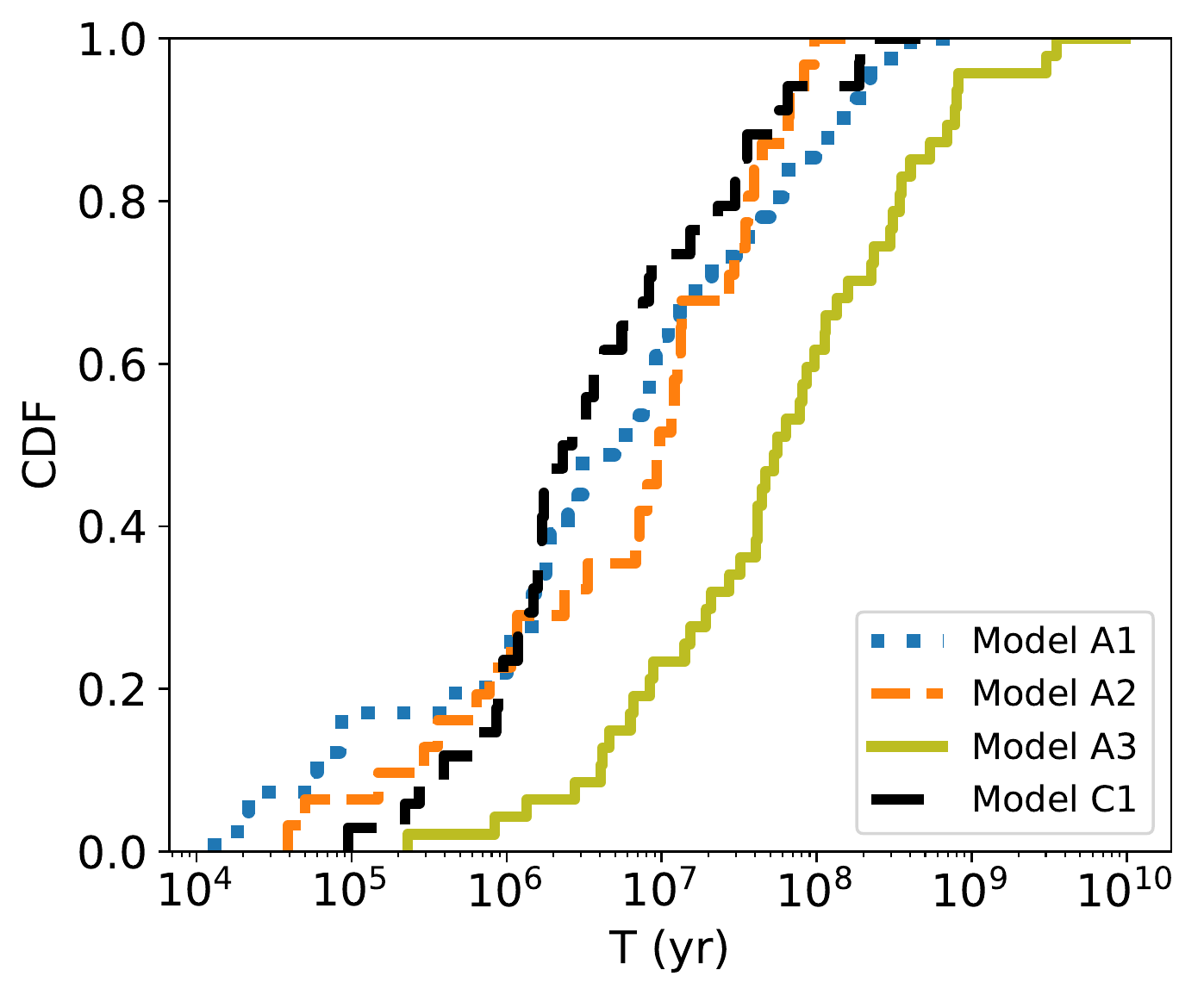}
\hspace{0.5cm}
\includegraphics[scale=0.55]{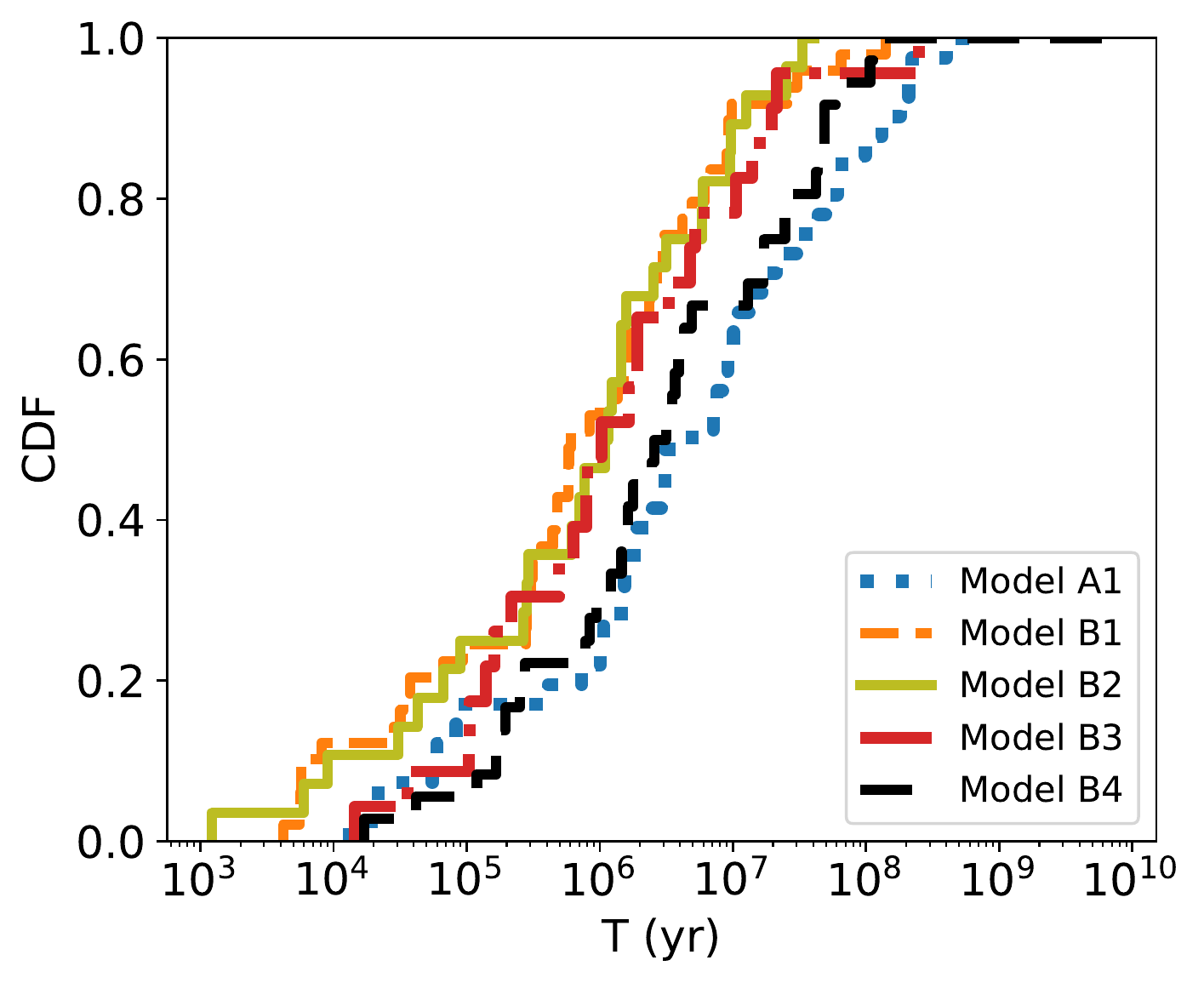}
\caption{Merger time distribution (after the SN events) of BH-NS binaries in triples that lead to merger (see Table~\ref{tab:models}). Left panel: different values of $\sigma$ and $\amax$; right panel: different values of $Z$.}
\label{fig:tmerge}
\end{figure*}

Figure~\ref{fig:tmerge} shows the merger time (after the SN events) CDFs of BH-NS binaries in triples that lead to merger for all models. The CDF does not depend significantly on $Z$, but depends mostly on $\sigma$. Larger kick velocities imply a larger outer semi-major axis, thus a larger typical KL timescale since $T_{\rm}\propto a_{\rm out,n}^{3}$. Compared to the case of $\sigma=0\kms$, the typical merger time is $2$--$3$ times longer when $\sigma=260\kms$. Different $\amax$'s (Model C1) do not affect significantly the merger time distribution.

In order to compute the merger rate of BH-NS binaries, we follow the scheme adopted in \citet{frl2019}. Assuming a local star formation rate of $0.025 \msun$ Mpc$^{-3}$ yr$^{-1}$, the number of stars formed per unit mass is given by \citep{both2011},
\begin{equation}
N(m)dm=5.4\times 10^6 m^{-2.3}\ \mathrm{Gpc}^{-3}\ \mathrm{yr}^{-1}\ ,
\end{equation}
and assuming a constant star-formation rate per comoving volume unit, the merger rate of binary BH-NS in triples is,
\begin{equation}
\Gamma_\mathrm{BH-NS}=8.1\times 10^4 f_{\rm 3} f_{\rm stable} f_{\rm merge}\ \mathrm{Gpc}^{-3}\ \mathrm{yr}^{-1}\ .
\label{eqn:rrr}
\end{equation}
Here, $f_{\rm 3}$ is the fraction massive stars in triples, $f_{\rm stable}$ is the fraction of triple systems with an inner BH-NS binary that remain stable after the all the SN events take place, and $f_{\rm merge}$ is the conditional probability that systems that are stable after all the SN events merge as a consequence of the KL mechanism. We adopt $f_{\rm 3}=0.15$ in our calculations. We find that the fraction of stable systems depends both on $\sigma$ and on the progenitor metallicity, since lower $Z$'s produce more massive BHs, that on average receive lower kicks as a result of our assumption of momentum-conserving kicks. We find $f_{\rm stable}\approx 1.4\times 10^{-2}$, $1.8\times 10^{-5}$, $2.6\times 10^{-7}$ for $\sigma=0\kms$, $100\kms$, $260\kms$, respectively, when $\amax=2000$ AU and $Z=0.01$, and $f_{\rm stable}\approx 5.5\times 10^{-4}$, $1.4\times 10^{-4}$, $1.6\times 10^{-6}$, $2.6\times 10^{-7}$, $5.1\times 10^{-8}$, for $Z=0.0001$, $Z=0.001$, $Z=0.005$, $Z=0.01$, $Z=0.015$, respectively, when $\sigma=260\kms$ and $\amax=2000$ AU. In the case $\amax=5000$ AU, we find that the fraction of stable systems is about half of the case where $\amax=2000$ AU. Finally, we find that the typical fraction of systems that merge is $f_{\rm merge}\sim 0.1$ (see Table~\ref{tab:models}). Plugging these numbers into Eq.~\ref{eqn:rrr},
\begin{equation}
\Gamma_\mathrm{BH-NS}=1.9\times 10^{-4}-22 \ \mathrm{Gpc}^{-3}\ \mathrm{yr}^{-1}\ .
\end{equation}
Our estimate for the rate overlaps with BH-NS mergers in binaries \citep{kruc2018} and is entirely within the LIGO allowed values \citep{ligo2018}.

\section{Conclusions}
\label{sect:conc}

The possible recent detection of a BH-NS merger event by the LIGO-Virgo collaboration via GWs attracted much attention to these sources. BH-NS mergers are of high interest for their possible EM counterparts, such as short gamma-ray bursts, which can observed by high-energy observatories and could provide crucial information on the related physics. While the formation of BH-NS binaries is not straightforward in star clusters as a result of the strong heating by BHs, isolated binary and triple systems can produce this type of mergers.

In this second paper of the series, we have followed-up our initial study \citep{frl2019} of the dynamical evolution of triples comprised of an inner BH-NS binary. We have focused on how the progenitor metallicity affects the distributions of the relevant parameters of the BH-NS binaries that undergo a merger in triples. We have determined the distributions of BH masses, orbital parameters, and merger times as a function of the progenitor metallicity and initial triple orbital distributions, and derive a merger rate range of $\Gamma_\mathrm{BH-NS}=1.9\times 10^{-4}-22 \ \mathrm{Gpc}^{-3}\ \mathrm{yr}^{-1}$. Our range is in agreement with the rate derived from isolated binaries and the LIGO upper limit. The dependence of the BH masses, orbital parameters and rates on the metallicity could be tested by either identifying the host galaxy from EM counterpart or by detecting the evolution of the source properties with increasing redshift, thus decreasing metallicity on average.

We have also shown that the typical eccentricity of BH-NS binaries that merge in triple systems in the LIGO band is $\sim 10^{-2}-10^{-1}$. Recent studies have shown that the orbital eccentricity is a useful tool for disentangling different production channels. Since GW emission is highly efficient at circularizing the orbit of an inspiraling BH binary, BHs that merge in isolation are expected to enter the LIGO frequency band ($10$ Hz) with very low eccentricities, $\sim 10^{-7}$--$10^{-6}$. In case the BH binary merger is catalyzed by KL cycles in a hierarchical system, as in our case, a number of authors showed that the typical eccentricity distribution peaks at much higher values, $\sim 10^{-2}$--$10^{-1}$, for various astrophysical scenarios \citep{antchrod2016,frbr2019,fragk2019,frl2019}. The spectrum of eccentricities at LIGO band is much richer for BHs merging as a result of the dynamical assembly in a star cluster \citep{sams2014,ssdo2018}. \citet{zevin18} demonstrated that there are three different possibilities in this scenario: (i) binaries that are ejected and merge outside the cluster have eccentricities $\sim 10^{-7}$--$10^{-6}$, as in the isolated binary case; (ii) binaries that merge as a result of a GW capture process have eccentricities $\sim 10^{-2}$--$10^{-1}$, as in the KL-induced mergers; (iii) binaries that merge within the cluster have intermediate eccentricities $\sim 10^{-5}$--$10^{-3}$. In any case, matched-filtering searches for GWs do not utilize eccentric templates, thus no eccentric mergers have been observed so far by the LIGO-Virgo network.

BH and NS spin magnitudes can be a powerful observables to constrain the physics of massive stars. Spin magnitudes are expected to be set by the physics governing the stellar collapse, which can depend on the progenitor star metallicity and mass loss. How the specific properties of the progenitor set the properties of the remnant spin is still highly uncertain. Therefore, we have also investigated the expected spin-orbit misalignments of merging BH-NS binaries. In this channel, the spins of the BHs and NSs in the inner binary can undergo a relativistic precession around the inner angular momentum. Using a quadrupole approximation, \citet{antonini2018} and \citet{rodant2018} argued that BH binaries merging in a triple system would lead typically to near-zero effective spins. However, more recently, \citet{liulai2019} have showed by using a more accurate integration of the equations of motions that this is not the case and the effective spins are rather distributed more uniformly. We have found that typically the effective spin distribution is peaked at $\chieff\sim 0$, but with significant tails. Regarding other scenarios, binaries that evolve and merge in isolation are expected to have spin vectors that are relatively aligned with the angular momentum of the binary, thus having an effective spin always positive with $\cos\thetabh$,$\cos\thetans\sim 1$ \citep{vitale2017,gerosa2018}. On the other hand, dynamically-assembled BHs in star clusters are expected to have spin vectors distributed isotropically with respect to the orbital angular momentum, thus leading to a distribution peaked at $\chieff\sim 0$ with symmetric tails \citep{rodze2016,arcaben2019}.

We note that in our simulations we check that the triples are stable after each SN event. However, systems that become unstable as a result of a SN may still produce a BH-NS merger, which we do not take into account in our results. We also note that we are assuming that the SNe take place instantaneously and do not simulate the systems during the main sequence lifetime of the progenitors. This and the details of the specific progenitor evolutions, which depend on winds, metallicity and rotation, could reduce the available parameter space for BH-NS mergers \citep[see e.g.][]{shapp2013}. The situation is even more complicated if mass loss during possible episodes of Roche-lobe overflows and common evolution phases in the triple are taken into account, which are not modeled as accurately as in binary systems \citep{rosa2019,hamd2019}. Mass transfer may likely harden the triple systems, possibly reducing the KL timescale thus the merger time. We also did not account for the fact that there could be an excess of systems with near-unity mass ratios and that the multiple fraction may be higher at low metallicities, if the triple were to follow the same trend of binaries \citep{moe2017,moe2019}. A near-unity mass ratio for the outer binary in our triples would imply a larger number of systems that remain bound after SNe take place, thus enhancing the merger rate of BH-NS binaries in triples.

The observation via GW emission of a merging BH-NS binary which enters the LIGO band with a high eccentricity and with a nearly zero effective spin would be a strong signature that the mechanism proposed in this paper is at work.

\section*{Acknowledgements}

GF thanks Rosalba Perna and Raffaella Schneider for useful discussions on stellar evolution, and Johan Samsing for comments on an earlier version of the manuscript. GF thanks Seppo Mikkola for helpful discussions on the use of the code \textsc{archain}. GF acknowledges support from a CIERA postdoctoral fellowship at Northwestern University. This work was also supported by the Black Hole Initiative at Harvard University, which is founded by a JTF grant (to AL).

\bibliographystyle{mn2e}
\bibliography{refs}

\end{document}